\documentclass[12pt]{article}
\usepackage[latin9]{inputenc}
\usepackage{geometry}
\usepackage{float}
\usepackage{amsmath}
\usepackage{amsthm}
\usepackage{amssymb}
\usepackage{graphicx}
\usepackage{setspace}
\usepackage{comment}

\usepackage[authoryear]{natbib}
\makeatletter

\providecommand{\tabularnewline}{\\}
\floatstyle{ruled}
\newfloat{algorithm}{tbp}{loa}
\providecommand{\algorithmname}{Algorithm}
\floatname{algorithm}{\protect\algorithmname}

\usepackage{latexsym}
\usepackage{amsthm}\usepackage{amsfonts}\usepackage{graphicx}\usepackage{epsfig}
\usepackage{bm}
\newcommand*{\patchAmsMathEnvironmentForLineno}[1]{%
      \expandafter\let\csname old#1\expandafter\endcsname\csname #1\endcsname
      \expandafter\let\csname oldend#1\expandafter\endcsname\csname end#1\endcsname
      \renewenvironment{#1}%
         {\linenomath\csname old#1\endcsname}%
         {\csname oldend#1\endcsname\endlinenomath}}%
    \newcommand*{\patchBothAmsMathEnvironmentsForLineno}[1]{%
      \patchAmsMathEnvironmentForLineno{#1}%
      \patchAmsMathEnvironmentForLineno{#1*}}%
    \AtBeginDocument{%
    \patchBothAmsMathEnvironmentsForLineno{equation}%
    \patchBothAmsMathEnvironmentsForLineno{align}%
    \patchBothAmsMathEnvironmentsForLineno{flalign}%
    \patchBothAmsMathEnvironmentsForLineno{alignat}%
    \patchBothAmsMathEnvironmentsForLineno{gather}%
    \patchBothAmsMathEnvironmentsForLineno{multline}%
    }
\usepackage[mathlines,displaymath]{lineno}

\include{def}
\def\dispmuskip{\thinmuskip= 3mu plus 0mu minus 2mu \medmuskip=  4mu plus 2mu minus 2mu \thickmuskip=5mu plus 5mu minus 2mu}
\def\textmuskip{\thinmuskip= 0mu                    \medmuskip=  1mu plus 1mu minus 1mu \thickmuskip=2mu plus 3mu minus 1mu}
\def\beq{\dispmuskip\begin{equation}}    \def\eeq{\end{equation}\textmuskip}
\def\beqn{\dispmuskip\begin{displaymath}}\def\eeqn{\end{displaymath}\textmuskip}
\def\bea{\dispmuskip\begin{eqnarray}}    \def\eea{\end{eqnarray}\textmuskip}
\def\bean{\dispmuskip\begin{eqnarray*}}  \def\eean{\end{eqnarray*}\textmuskip}

\newcommand{\eps}{\epsilon}
\newcommand{\wh}{\widehat}

\def\E{{\mathbb E}}                         
\def\V{{\mathbb V}}

\usepackage{xr}
\makeatletter
\newcommand*{\addFileDependency}[1]{
  \typeout{(#1)}
  \@addtofilelist{#1}
  \IfFileExists{#1}{}{\typeout{No file #1.}}
}
\makeatother

\newcommand*{\myexternaldocument}[1]{
    \externaldocument{#1}
    \addFileDependency{#1.tex}
    \addFileDependency{#1.aux}
}

\myexternaldocument{OnlineSupplement}
\myexternaldocument{reply2refereesJCGS_R2}

\usepackage[mathlines,displaymath]{lineno}

\def\dlq{\lq \lq}
\def\drq{\rq \rq}

\def\Eq{Eq. \eqref}

\usepackage{authblk}
\makeatother
\begin{document}
\title{Flexible Variational Bayes based on a Copula of a Mixture}

\author[1,3]{David Gunawan}
\author[2,3]{Robert Kohn}
\author[4,5]{David Nott}
\affil[1]{School of Mathematics and Applied Statistics, University of Wollongong}
\affil[2]{School of Economics, UNSW Business School, University of New South Wales}
\affil[3]{Australian Center of Excellence for Mathematical and Statistical Frontiers}
\affil[4]{Department of Statistics and Data Science, National University of Singapore}
\affil[5]{Institute of Operations Research and Analytics, National University of Singapore}
\renewcommand\Authands{ and }
\maketitle

\begin{abstract}
Variational Bayes methods approximate the posterior density by a family
of tractable distributions whose parameters are estimated by optimisation. Variational approximation is useful
when exact inference is intractable or very costly.
Our article develops a flexible variational approximation based on a copula
of a mixture, which is implemented by combining boosting, natural gradient, and a variance reduction method.
The efficacy of the approach is illustrated by using simulated and real datasets
to approximate  multimodal, skewed and
heavy-tailed posterior distributions, including an application to
Bayesian deep feedforward neural network regression models. Supplementary materials, including appendices and computer code for this article, are available online. 
\bigskip{}

\end{abstract}


\section{Introduction \label{sec:Introduction}}
Variational Bayes (VB)  methods are increasingly used for
Bayesian inference in a wide range of challenging statistical
models~\citep{Ormerod:2010,Blei2017}. VB approximates the target posterior density as the solution of an optimisation problem over a simpler and more tractable family of distributions; this family
is usually selected to balance accuracy and computational cost. 
VB methods are
particularly useful in estimating the posterior densities of the parameters
of complex statistical models when
exact inference is impossible or computationally expensive.
They are usually computationally much cheaper than methods such as Markov chain Monte Carlo (MCMC) which produce exact draws from the posterior as the number of simulated draws goes to infinity. We call this property of MCMC estimators \lq simulation consistent\rq{}, and for the rest of the paper, we refer to MCMC-type algorithms as \lq exact\rq{}  methods. 


Much of the current literature focuses on Gaussian variational approximation (GVA)
for approximating the target posterior density \citep{Challis2013,Titsias2014,Kucukelbir2017,Tan2018,Ong2018}.
A major problem with Gaussian approximations is that many posterior distributions are 
skewed, multimodal,
and heavy-tailed. This is true in particular for complex statistical models such as Bayesian deep feedforward neural
network (DFNN) regression models
\citep{jospin2022hands,Izmailov2021}.
Gaussian variational approximations for such models
poorly approximate their posterior distributions; see section \ref{sec:Examples}.

There are a number of attempts to overcome the issue with Gaussian variational approximation including \citet{Smith2020} who 
propose Gaussian copula and skew Gaussian copula-based variational approximations;  \citet{Guo2016}  and \citet{Miller2016} who propose a mixture of normals variational approximation; \citet{rezende2015variational} who propose normalizing planar and radial flows. Other types of normalising flows for variational inference are also proposed in the literature and are reviewed by \citet{papamakarios2021normalizing}.

Our article makes a number of contributions. The {\sf first} is to 
propose a flexible copula-based variational approximation by a mixture of Gaussians
that builds  on the Gaussian and  
skew-Gaussian copula approximations \citep{Smith2020} and on the 
mixture of Gaussians variational approximations \citep{Guo2016,Miller2016}. We do not believe that such a variational mixture has been used before in the literature. 
The main idea of this part of our approach is to start by using a  Gaussian or skew Gaussian copula-based approximation as the first component. This leads to a possible transformation of each parameter (marginal of the posterior) which may then make it easier to fit a joint distribution to the posterior of the {\em transformed} parameters. 
In our approach simpler Gaussian distributions are then used as the additional mixture components to make the optimisation tractable while still improving the inference and prediction.
Section \ref{sec:Variational-Approximation} gives further details. The key insight is 
that of fitting a mixture after transforming each parameter, rather than using a simple multivariate family like the normal distribution. 
The proposed variational approximation allows fitting of multimodal, skewed, heavy tailed, and high-dimensional
posterior distributions with complex dependence structures. We show
in a number of examples that our approach gives more accurate inference
and forecasts than the corresponding Gaussian copula and mixture of normals variational approximation. Although not proved in the paper,  it is not difficult to see that a version of an estimator based on a  copula of a mixture  will be a universal approximator of a multivariate distribution under reasonable assumptions because a mixture of normals is a universal approximator. It is interesting to note that a mixture of Gaussian copulas does not give a universal approximator of a multivariate distribution \citep{Khaled2016}. 


Variational optimisation of a 
copula-based mixture approximation
is challenging in complex models with a large number of parameters
because of the large
number of variational parameters that need to be optimised.
The boosting variational inference method in \citet{Guo2016},  and \citet{Miller2016}
is a promising recent approach to fit mixture type variational approximations.
By adding a single mixture component at a time, the posterior
approximation of the model parameters is refined iteratively. \citet{Miller2016}
uses stochastic
gradient ascent (SGA) optimisation with the reparameterisation trick to fit a
mixture of Gaussian densities. They find that the use of the reparameterisation trick in the boosting variational method still results in a large variance and it is necessary
 to use many samples to estimate the gradient of the variational lower bound.
\citet{Guo2016}, \citet{locatello+kgr18} and \citet{campbell2019universal} consider similar variational boosting mixture approximations, although they
use different approaches to optimise and specify the mixture components. \citet{jerfel2021variational} consider boosting using the forward KL-divergence and combines variational inference and importance sampling. 

Our {\sf second} contribution is to build 
(see section~\ref{sec:Variational-Methods}) 
on the variational boosting
method by efficiently adding a single mixture component at a time using the natural gradient
\citep{Amira:1998} and the variance reduction method of \citet{Ranganath:2014}
to fit the flexible copula of the mixture. 
Previous literature on variational boosting optimisation
does not use the natural gradient.
Many natural-gradient variational methods are available, e.g.,
\citet{Hoffman:2013}, \citet{Khan2017a}, and \citet{Lin2019}. This literature
shows  that using natural gradients enables faster convergence
than the traditional gradient-based methods because they exploit the
information geometry of the variational approximation; section \ref{efficiencyNaturalGradients} suggests that this is also true for our estimator. 

Our  {\sf  third} contribution (see section \ref{subsec:Initialising-a-New})  is to provide methods to initialise the variational parameters of the additional component in the mixtures.

The rest of the article is organised as follows.
Section~\ref{sec:Summary-of-VB approach}
gives the necessary background to the paper;  section \ref{sec:Variational-Approximation}
discusses the copula of the mixture variational approximation;  section \ref{sec:Variational-Methods} discusses
the variational optimisation algorithm that fits the copula of a mixture variational approximation; section~\ref{sec:Examples}
presents results from both simulated and real datasets; section \ref{sec:Conclusions}  concludes
with a discussion of our approach and results. This article has an online supplement containing additional technical details and  empirical results.


\section{Variational Inference \label{sec:Summary-of-VB approach}}

Let $\theta\in\Theta$ be the vector of model parameters, and
$y_{1:n}=\left(y_{1},...,y_{n}\right)$ the vector of observations. Bayesian
inference about  $\theta$ is based on the posterior
distribution
\begin{equation*}
\pi\left(\theta\right)=
p\left(\theta|y_{1:n}\right)=\frac{p\left(y_{1:n}|\theta\right)
p\left(\theta\right)}{p\left(y_{1:n}\right)};
\end{equation*}
$p\left(\theta\right)$ is the prior, $p\left(y_{1:n}|\theta\right)$ is
the likelihood function, and $p\left(y_{1:n}\right)$ is the marginal
likelihood. The posterior distribution $\pi\left(\theta\right)$ is
unknown for most statistical models, making it challenging to
carry out Bayesian inference. We consider variational inference methods,
where  a member $q_{\lambda}\left(\theta\right)$ of some family
of tractable densities, indexed by the variational parameter $\lambda\in\Lambda$,
is used to approximate the posterior $\pi\left(\theta\right)$. The
optimal variational parameter $\lambda$ is chosen by minimising the
Kullback-Leibler divergence between $q_{\lambda}\left(\theta\right)$
and $\pi\left(\theta\right)$,
\begin{eqnarray*}
\textrm{KL}\left(\lambda\right) & := & \int\log\left ( \frac{q_{\lambda}\left(\theta\right)}{\pi\left(\theta\right)}\right )
q_{\lambda}\left(\theta\right)d\theta.\\
 & = & \int q_{\lambda}\left(\theta\right)\log\left ( \frac{q_{\lambda}\left(\theta\right)}{p\left(y_{1:n}|\theta\right)p\left(\theta\right)}\right ) d\theta+\log p\left(y_{1:n}\right)\\
 & = & -\textrm{\ensuremath{\mathcal{L}}}\left(\lambda\right)+\log p\left(y_{1:n}\right),
\end{eqnarray*}
where
\begin{equation}
\mathcal{L}\left(\lambda\right) :=\int\log\left ( \frac{{p\left(y_{1:n}|\theta\right)p\left(\theta\right)}}{q_{\lambda}\left(\theta\right)}\right ) q_{\lambda}\left(\theta\right)d\theta,\label{eq:LB}
\end{equation}
is a lower bound on the log of the marginal likelihood $\log p\left(y_{1:n}\right)$.
 Therefore, minimising the KL divergence between $q_{\lambda}\left(\theta\right)$
and $\pi\left(\theta\right)$ is equivalent to maximising the Evidence Lower Bound (ELBO) given by \Eq{eq:LB}. The ELBO can be used as a tool for model selection  \citep{Smith2020,Tran2020,Ong2018} although care is needed
if the tightness of the lower bound varies significantly between
the candidate models.


Although $\mathcal{L}\left(\lambda\right)$ is often an  intractable integral with no closed
 form solution, we can write it
as an expectation with respect to $q_{\lambda}\left(\theta\right)$,
\begin{equation}
\mathcal{L}\left(\lambda\right)=E_{q_{\lambda}}\left(\log g\left(\theta\right)-\log q_{\lambda}\left(\theta\right)\right),\label{eq:LB2}
\end{equation}
where $g\left(\theta\right) := p\left(y_{1:n}|\theta\right)p\left(\theta\right)$.
This interpretation 
allows the use
of  stochastic gradient ascent (SGA)  methods
to maximise the variational lower bound $\mathcal{L}\left(\lambda\right)$.
See, for e.g., \cite{Nott:2012}, \cite{Paisley:2012}, \cite{Hoffman:2013}, \cite{Salimans:2013}, \cite{Kingma2014},  
 \cite{Titsias2014}, and \cite{Rezende2014}.
In SGA, an initial value $\lambda^{\left(0\right)}$ is updated according
to the iterative scheme
\begin{equation}
\lambda^{\left(t+1\right)} := \lambda^{\left(t\right)}+a_{t}\circ\widehat{\nabla_{\lambda}\mathcal{L}\left(\lambda^{\left(t\right)}\right)},
\; \textrm{  for } t=0,1,2,... ; \label{eq:stochastic gradient update}
\end{equation}
$\circ$ denotes the Hadamard (element by element) product of
two random vectors; $a_{t}:=\left(a_{t1},...,a_{tm}\right)^{\top}$
is a vector of step sizes, where
 $m$ is the dimension of  the variational parameters 
$\lambda$, and $\widehat{\nabla_{\lambda}\mathcal{L}\left(\lambda^{\left(t\right)}\right)}$
is an unbiased estimate of the gradient of the lower bound $\mathcal{L}\left(\lambda\right)$
at $\lambda=\lambda^{\left(t\right)}$. The learning rate sequence
satisfies the Robbins-Monro conditions $\sum_{t}a_{t}=\infty$
and $\sum_{t}a_{t}^{2}<\infty$ \citep{Robbins1951}, which ensures
that the iterates $\lambda^{\left(t\right)}$ converge to a local
optimum as $t\rightarrow\infty$ under suitable regularity conditions
\citep{bottou2010}. Adaptive step sizes are often used in practice,
and we employ the ADAM method of \citet{Kingma2015}, which uses bias-corrected
estimates of the first and second moments of the stochastic gradients
to compute adaptive learning rates. The update in  \Eq{eq:stochastic gradient update}
continues until a stopping criterion is satisfied.

To estimate the gradient, SGA methods often use the \dlq log-derivative
trick\drq{} \citep{Kleijnen1996},  $\left(E_{q_{\lambda}}\left(\nabla_{\lambda}\log q_{\lambda}\left(\theta\right)\right)=0\right)$,
and it is straightforward to show that
\begin{equation}
\nabla_{\lambda}\mathcal{L}\left(\lambda\right)=E_{q_{\lambda}}\left(\nabla_{\lambda}\log q_{\lambda}\left(\theta\right)\left(\log g(\theta)-\log q_{\lambda}\left(\theta\right)\right)\right); \label{eq:standardgradientlowerbound}
\end{equation}
 $E_{q_{\lambda}}$ is the expectation with respect to $q_{\lambda}\left(\theta\right)$
in \Eq{eq:standardgradientlowerbound}.
Let
\begin{equation*}
g_{\lambda_{i}}:=\frac{1}{S}\sum_{s=1}^{S}\left(\log g(\theta_{s})-\log q_{\lambda}\left(\theta_{s}\right)\right)\nabla_{\lambda_{i}}\log q_{\lambda}\left(\theta_{s}\right).
\end{equation*}
Then,
 $\left(g_{\lambda_{1}},g_{\lambda_{2}},...,g_{\lambda_{m}}\right)^{\top} $
 is an unbiased estimate of $\nabla_{\lambda}\mathcal{L}\left(\lambda\right)$. However, this approach usually results in large fluctuations
in the stochastic gradients \citep{Ranganath:2014}. Section \ref{sec:Variational-Methods} discusses variance reduction and natural gradient
methods, which are very important for
fast convergence and stability.

\section{Flexible Variational Approximation based on a Copula of a Mixture
\label{sec:Variational-Approximation}}

\citet{Smith2020} propose Gaussian and skew Gaussian copula based variational approximations, which are constructed using Gaussian or skew Gaussian distributions after element-wise
 transformations of the parameters. They consider
the Yeo-Johnson \citep{Yeo2000} and G\&H families \citep{Tukey1977} for the
element-wise transformations and use the sparse factor structure
proposed by \citet{Ong2018} as the covariance matrix of the Gaussian
distributions. This section discusses the flexible copula based mixture of Gaussians variational approximation that builds
on  \citet{Smith2020}. The main idea is to use a flexible variational approximation, such as the Gaussian or skew Gaussian copulas of \citet{Smith2020} as the first component. This step also produces the transformed parameters we work with for the rest of the components. The rest of the components (second, third, etc.)  are then chosen to be Gaussian distributions with a simple
covariance structure making the optimisation tractable, while still improving the inference and prediction.


Let $t_{\gamma}(\theta) := \left (t_{\gamma_1} (\theta_1) , \dots, t_{\gamma_m} (\theta_m)\right )^\top$ be a family of one-to-one transformations with parameter
vector $\gamma$. Each parameter $\theta_i$ is first transformed as $\varphi_{i}:=t_{\gamma_{i}}\left(\theta_{i}\right)$;
the density of $\varphi:= \left ( \varphi_1, \dots, \varphi_m\right)^\top$ is then modeled as a $K$ component multivariate mixture of Gaussians.
The  variational approximation density for $\theta$ is then  obtained
by computing the Jacobian of the element-wise transformation from
$\theta_{i}$ to $\varphi_{i}$, for $i=1,...,m$, so that
\begin{equation}
q_{\lambda}\left(\theta\right) :=\sum_{k=1}^{K}\pi_{k}N\left(\varphi|\mu_{k},\Sigma_{k}\right)\prod_{i=1}^{m}\dot t_{\gamma_{i}}\left(\theta_{i}\right); \label{eq:thevariationalmixturesapproximation}
\end{equation}
the variational parameters are
\begin{equation*}
\lambda=\left(\gamma^{\top}_{1},...,\gamma^{\top}_{m},\left(\mu^{\top}_{1},...,\mu^{\top}_{K}
\right),\left(\pi_{1},...,\pi_{K}\right),\left(\textrm{vech}(\Sigma_{1})^{\top},...,\textrm{vech}
(\Sigma_{K})^{\top}\right)\right)^{\top}  \footnote{For an $m\times m$
matrix $A$, $\textrm{vec}\left(A\right)$ is the vector of length
$m^{2}$ obtained by stacking the columns of $A$ under each other
from left to right;  $\textrm{vech}\left(A\right)$ is the vector of length
$m\left(m+1\right)/2$ obtained from $\textrm{vec}\left(A\right)$
by removing all the superdiagonal elements of the matrix $A$.};
\end{equation*}
$\dot t_{\gamma_{i}}\left(\theta_{i}\right):=d\varphi_{i}/ d\theta_{i}$
and $\sum_{k=1}^{K}\pi_{k}=1$. The marginal densities of the approximation
are
\begin{equation}
q_{\lambda_{i}}\left(\theta\right)=\sum_{k=1}^{K}\pi_{k}N\left(\varphi_{i}|\mu_{k,i},\Sigma_{k,i}\right)
\prod_{i=1}^{m}\dot t_{\gamma_{i}}\left(\theta_{i}\right),\label{eq:marginalparameters}
\end{equation}
for $i=1,...,m$, with $\lambda_{i}=\left(\gamma^{\top}_{i},\left\{ \mu^{\top}_{k,i},\textrm{vech}(\Sigma_{k,i})^{\top}\right\} _{k=1}^{K},\left\{ \pi_{k}\right\} _{k=1}^{K}\right)^{\top}$,
a sub-vector of $\lambda$. As in \citet{Smith2020}, the variational
parameters $\lambda$ are all identified in $q_{\lambda}\left(\theta\right)$
without additional constraints because they are also parameters of
the margins given in Eq. \eqref{eq:marginalparameters}. 
The
variational approximation in Eq. \eqref{eq:thevariationalmixturesapproximation} is called a copula of a mixture of Gaussians variational approximation (CMGVA). It can fit multimodal, skewed, heavy tailed,
and high-dimensional posterior distributions with complex dependence
structures; see section \ref{sec:Examples}. 

When $\theta$ is high dimensional, we follow \citet{Ong2018}
and adopt a factor structure for each $\Sigma_{k}$, $k=1,...,K$. Let
$\beta_{k}$ be an $m\times r_{k}$ matrix, with $r_{k}\ll m$. For
identifiability, we set the strict upper triangle of $\beta_{k}$
to zero. Let $d_{k}=\left(d_{k,1},...,d_{k,m}\right)^{\top}$ be a
parameter vector with $d_{k,i}>0$, and denote by $D_{k}$ the $m\times m$
diagonal matrix with entries $d_{k}$. We assume that
\[
\Sigma_{k}:=\beta_{k}\beta_{k}^{\top}+D_{k}^{2},
\]
so that the number of parameters in $\Sigma_{k}$ grows linearly
with $m$ if $r_{k}\ll m$ is kept fixed. Note that the number of
factors $r_{k}$ can be different for each component of the mixture
for $k=1,...,K$. We set the number of factors for the first component higher than for the other components in the mixture to make the variational approximation scalable.
Section \ref{sec:Examples} discusses this further.

To draw $S$ samples from the variational approximation given in
Eq.~\eqref{eq:thevariationalmixturesapproximation},
the indicator variables $G_{s}$, for $s=1,...,S$, are first generated; each $G_s$
selects the
component of the mixture from which the sample is to be drawn, with
 $G_{s}=k$ with probability $\pi_{k}$. Then,
 $\left(z_{G_{s},s},\eta_{G_{s},s}\right)\sim N\left(0,I\right)$ are generated,
where $z_{G_{s},s}$ is $r_{G_{s}}$-dimensional and $\eta_{G_{s},s}$
is $m$-dimensional;  $\varphi_{s}=\mu_{G_{s}}+\beta_{G_{s}}z_{G_{s},s}+d_{G_{s}}\circ\eta_{G_{s},s}$ are then
calculated,
where $\circ$ denotes the Hadamard  product defined above. This representation shows that the latent
variables $z$, which are low-dimensional, explain all the correlation
between the transformed parameters $\varphi_{s}$, and the parameter-specific
idiosyncratic variance is captured by $\eta$. Finally,
$\theta_{s,i}=t_{\gamma_{i}}^{-1}\left(\varphi_{s,i}\right)$ is generated for
$i=1,...,m$ and $s=1,...,S$. The Yeo-Johnson (YJ) transformation \citep{Yeo2000} is used as $t_{\gamma_{i}}$ for all $i=1,...,m$.
If a parameter $\theta_i$ is constrained, it is first
transformed to the real line; for example, a variance parameter is transformed to its logarithm. \citet{Smith2020} gives the inverses and
derivatives of the Yeo-Johnson transformation. Both the Gaussian copula and the mixture of Gaussians variational approximations are special cases of the CMGVA. The mixture of Gaussians variational approximation is a special case of the  CMGVA when the variational parameters $\gamma_{i}$, are set to 1 
for $i=1,...,m$. The Gaussian copula is a special case of the CMGVA when the number of components in the mixture is $K=1$. 

The variational approximation in Eq. \eqref{eq:thevariationalmixturesapproximation} can be extended by including a skew Gaussian component as the first mixture component. It then becomes 

\begin{equation}
q_{\lambda}\left(\theta\right):=\left(\pi_{1}SN\left(\varphi|\mu_{1},\Sigma_{1},\widetilde{\alpha}_{1}\right)+\sum_{k=2}^{K}\pi_{k}N\left(\varphi|\mu_{k},\Sigma_{k}\right)\right)\prod_{i=1}^{m}\dot{t}_{\gamma_{i}}\left(\theta_{i}\right),
\label{eq:skewnormalva}
\end{equation}
where $SN\left(\varphi|\mu_{1},\Sigma_{1},\widetilde{\alpha}_{1}\right)$
is a multivariate skew Gaussian distribution of \citet{azzalini85} with
density 
\[
SN\left(\varphi|\mu_{1},\Sigma_{1},\widetilde{\alpha}_{1}\right)=2N\left(\varphi|\mu_{1},\Sigma_{1}\right)\Phi_{1}\left(\widetilde{\alpha}_{1}^{\intercal}S_{1}^{-1/2}\left(\varphi-\mu_{1}\right)\right),
\]
$S_{1}=\textrm{diag}\left(\sigma_{1}^{2},...,\sigma_{m}^{2}\right)$,
$\sigma_{i}^{2}$ is the $i$th diagonal element of $\Sigma_{1}$,
and $\widetilde{\alpha}_{1}=\left(\widetilde{\alpha}_{1,1},...,\widetilde{\alpha}_{1,m}\right)^{\top}$.
The parameter $\widetilde{\alpha}_{i,1}$ determines the level of
skewness of the $i$th marginal of $\varphi$. When $\widetilde{\alpha}_{i,1}=0$
for $i=1,...,m$, the distribution reduces to a multivariate Gaussian. The skew Gaussian copula is a special case of the variational approximation in Eq. \eqref{eq:skewnormalva} when the number of components in the mixture is $K=1$.


\section{Variational Methods\label{sec:Variational-Methods}}

This section discusses the estimation method for the copula based  mixture of Gaussians variational
approximation in 
Eq.~\eqref{eq:thevariationalmixturesapproximation}.
We first describe how the first component of the
mixture is fitted and then the process for adding an additional component to
the existing mixture approximation. Extension to the variational approximation in Eq. \eqref{eq:skewnormalva} is straightforward.

\subsection{Optimisation Methods\label{subsec:Optimisation-Methods}}

The method starts by fitting an approximation to the posterior distribution
$\pi\left(\theta\right)$ with a single mixture distribution, $K=1$, using the variational optimisation algorithm
given in \citet{Smith2020};  the optimal first component variational
parameters are denoted as $\lambda_{1}^{*}:=\left(\mu^{\top}_{1},\beta^{\top}_{1},d^{\top}_{1},\pi_{1},\gamma^{\top}\right)^{\top}$, with
the mixture weight $\pi_{1}$  set to $1$. We
do this by maximising the first lower bound objective function
\begin{align*}
\mathcal{L}^{\left(1\right)}\left(\lambda_{1}\right) & =  E_{q_{\lambda}}\left(\log g\left(\theta\right)-\log q_{\lambda_{1}}^{\left(1\right)}\left(\theta\right)\right),\quad
\lambda_{1}^{*}  =  \underset{\lambda_{1}}{\textrm{arg max}}\; \mathcal{L}^{\left(1\right)}\left(\lambda_{1}\right),
\end{align*}
where $q_{\lambda_{1}}^{\left(1\right)}\left(\theta\right)=N\left(\varphi;\mu_{1},\beta_{1}\beta_{1}^{\top}+D_{1}^{2}\right)\prod_{i=1}^{m}\dot t_{\gamma_{i}}\left(\theta_{i}\right)$.  After the optimisation algorithm converges,  $\lambda_{1}$ is fixed as
$\lambda_{1}^{*}$.



%

After iteration $K$, the current approximation to the posterior distribution
$\pi\left(\theta\right)$ is a mixture distribution with $K$ components
\begin{equation*}
q_{\lambda}^{\left(K\right)}\left(\theta\right)=
\sum_{k=1}^{K}\pi_{k}N\left(\varphi|\mu_{k},\beta_{k}\beta_{k}^{\top}+
D_{k}^{2}\right)\prod_{i=1}^{m}\dot t_{\gamma_{i}}\left(\theta_{i}\right).
\end{equation*}
We can introduce a new mixture component with new component parameters,
$\left(\mu_{K+1},\beta_{K+1},d_{K+1}\right)$, and a new mixing weight
$\pi_{K+1}$. The weight $\pi_{K+1}\in\left[0,1\right]$ mixes
between the new component and the existing approximation. The new
approximating distribution is
\begin{multline*}
q_{\lambda}^{\left(K+1\right)}\left(\theta\right)
:=\left(\left(1-\pi_{K+1}\right)\left(\sum_{k=1}^{K}
\pi_{k}N\left(\varphi|\mu_{k},\beta_{k}\beta_{k}^{\top}+D_{k}^{2}\right)\right)+\right.\\
\left.\pi_{K+1}N\left(\varphi|\mu_{K+1},\beta_{K+1}\beta_{K+1}^{\top}+D_{K+1}^{2}
\right)\right)\prod_{i=1}^{m}\dot t_{\gamma_{i}}\left(\theta_{i}\right).
\end{multline*}
The new lower bound objective function is
\begin{eqnarray*}
\mathcal{L}^{\left(K+1\right)}\left(\lambda_{K+1}\right) & := & E_{q_{\lambda}}\left(\log g\left(\theta\right)-\log q_{\lambda}^{\left(K+1\right)}\left(\theta\right)\right),\\
\lambda_{K+1}^{*} & := & \underset{\lambda_{K+1}}{\textrm{arg max}}\; \mathcal{L}^{\left(K+1\right)}\left(\lambda_{K+1}\right).
\end{eqnarray*}

Note that with the copula transformation fixed, updating the mixture approximation parameters is the same as updating an ordinary mixture approximation in the transformed space of $\varphi$.
Since the existing variational approximation is also fixed, it is only necessary
to optimise the new component parameters $\left(\mu_{K+1},\beta_{K+1},d_{K+1}\right)$,
and the new mixing weight $\pi_{K+1}$, which reduces the dimension
of the variational parameters to be optimised. Although the existing mixture components are fixed, their mixing weights can vary.
It is possible to reoptimise the variational parameters $\gamma_{i}$ for all $i=1,...,m$ at each iteration of the algorithm. However, we obtained  no substantial improvement
with the increased computational cost. The $\gamma_{i}$, for all $i=1,...,m,$ are kept fixed for iterations $k>1$. 



\subsection{Updating the Variational Parameters\label{subsec:Updating-the-Variational}}

This section outlines the updating scheme for the variational parameters
of the new component parameters $\left(\mu_{K+1},\beta_{K+1},D_{K+1}\right)$
and the new mixing weight $\pi_{K+1}$ based on natural-gradient
methods and control-variates for reducing the variance of the unbiased
estimates of the gradient of the lower bound. Many natural-gradient
methods for variational inference are available
\citep{Hoffman:2013,Khan2017a}; these
show that natural-gradients produce faster convergence
than traditional gradient-based methods.


The natural-gradient approach exploits the information geometry of
the variational approximation $q_{\lambda}{(\theta)}$ to speed-up the convergence
of the optimisation. If the Fisher information matrix (FIM),
denoted by $F_{\lambda}$, of the $q_{\lambda}\left(\theta\right)$
is positive-definite for all $\lambda\in\Lambda$, the natural-gradient
update is
\begin{equation}
\lambda^{\left(t+1\right)}=\lambda^{\left(t\right)}+a_{t}\circ\left(F_{\lambda}^{-1}
\widehat{\nabla_{\lambda}\mathcal{L}\left(\lambda^{\left(t\right)}\right)}\right),\;\;\textrm{for } t=1,2,...\label{eq:NG}
\end{equation}
Multiplying the estimated gradient of the
lower bound by the inverse of the Fisher information matrix leads to a proper scaling of the gradient in each dimension
and takes into account dependence between the variational parameters
$\lambda$. In general, the natural-gradient update in 
Eq.~\eqref{eq:NG}
requires computing and inverting the FIM, which can be computationally
expensive in high-dimensional problems. However,  \citet{Khan2018} shows
that for exponential families (EF), the natural-gradient
update can be much simpler than the traditional gradient-based methods.
 The standard EF variational approximation is
\begin{equation*}
q_{\lambda}\left(\theta\right)=h\left(\theta\right)\exp\left[\left\langle \phi\left(\theta\right),\lambda\right\rangle -A\left(\lambda\right)\right],
\end{equation*}
where $\phi\left(\theta\right)$ is the sufficient statistic, $h\left(\theta\right)$
is the base measure, $A\left(\lambda\right)$ is the log-partition, and $\langle\cdot,\cdot\rangle$ denotes an inner product.
For such approximations, it is unnecessary to compute
the Fisher information matrix (FIM)  explicitly and the expectation parameter $m_{\theta}\left(\lambda\right)=\E_{q}\left(\phi\left(\theta\right)\right)$
can be used to compute natural-gradients, provided the FIM is invertible for all $\lambda$.
The  update for the natural-gradient method is now
\begin{equation}
\lambda^{\left(t+1\right)}=\lambda^{\left(t\right)}+a_{t}\circ\left(\widehat{\nabla_{m_{\theta}}
\mathcal{L}\left(\lambda^{\left(t\right)}\right)}\right),\;\; \textrm{for }t=1,2,...\label{eq:NG-1}
\end{equation}
where $\nabla_{m_{\theta}}$ denotes the gradient with respect to the expectation parameter ${m_{\theta}}$. The following  is used to obtain
Eq.~\eqref{eq:NG-1}:
\begin{equation*}
\nabla_{\lambda}\mathcal{L}\left(\lambda\right)=
\left[\nabla_{\lambda}m_{\theta}^{\top}\right]\nabla_{m_{\theta}}\mathcal{L}\left(\lambda\right)=
\left[F_{\lambda}\right]\nabla_{m_{\theta}}\mathcal{L}\left(\lambda\right);
\end{equation*}
the first equality is obtained by using the chain rule
and the second equality is obtained by noting $\nabla_{\lambda}m_{\theta}^{\top}=\nabla_{\lambda}^{2}A\left(\lambda\right)=F_{\lambda}$;
see \citet{Lin2019} for details.


\citet{Lin2019} derive natural gradient methods for a mixture of EF distributions in the conditional exponential family form
\begin{equation}
q_{\lambda}\left(\theta,\underline{w}\right)
=q_{\lambda}\left(\theta|\underline{w}\right)q_{\lambda}\left(\underline{w}\right),
\end{equation}
with $q_{\lambda}\left(\theta|\underline{w}\right)$ as the component
and $q_{\lambda}\left(\underline{w}\right)$ as the mixing distribution.
A special case is the finite mixture of Gaussians, where the components
in EF form are mixed using a multinomial distribution. They show that
if the FIM, $F_{\lambda}\left(\theta,\underline{w}\right)$,
of the joint distribution of $\theta$ and $\underline{w}$, is invertible, then
it is possible to derive natural-gradient updates without explicitly
computing the FIM. They use the update
\begin{equation*}
\lambda^{\left(t+1\right)}:=\lambda^{\left(t\right)}+a_{t}
\circ\left(\widehat{\nabla_{m}\mathcal{L}\left(\lambda^{\left(t\right)}\right)}\right),\;\; \textrm{for }t=1,2,...,
\end{equation*}
with the expectation parameters $m:=\left(m_{\theta},m_{\underline{w}}\right)$,
where $m_{\theta}:=E_{q_{\lambda}\left(\theta|\underline{w}\right)
q_{\lambda}\left(\underline{w}\right)}\left(\phi\left(\theta,\underline{w}\right)\right)$,
with
 $\phi\left(\theta,\underline{w}\right):=\left\{ \textrm{I}_{k}\left(\underline{w}\right)\theta,\textrm{I}_{k}\left(\underline{w}\right)
\theta\theta^{\top}\right\} _{k=1}^{K-1}$, and $m_{\underline{w}}:=E_{q_{\lambda}\left(\underline{w}\right)}\left(\phi\left(\underline{w}\right)\right)$ with $\phi(\underline{w})=\left\{ \textrm{I}_{k}\left(\underline{w}\right)\right\} _{k=1}^{K-1}$, and $\textrm{I}_k(\underline{w})$ denotes the indicator function which is 1 if $\underline{w}=k$, and $0$ otherwise.
This results in simple natural-gradient updates for the mixture
components and weights. \citet{Lin2019} apply their natural gradient methods for fitting a mixture of Gaussians variational approximation with a full covariance matrix for each component, which makes it less scalable for a large number of parameters. They also do not use the boosting approach for adding a mixture component one at a time. In this case, choosing good initial values for all variational parameters can be difficult.  

As  a factor structure is used for the covariance
matrix, we adopt the natural-gradient updates of \citet{Lin2019} only for the new weight
$\pi_{K+1}$ and the mixture means $\mu_{K+1}$. Denote $\pi^\prime_{1}:=\left(1-\pi_{K+1}\right)\sum_{k=1}^{K}\pi_{k}$
and $\pi^\prime_{2}:=\pi_{K+1}$, $\pi^\prime_{1}+\pi^\prime_{2}=1$. The natural-gradient
update for the new mixture weight, $\pi_{K+1}$ is
\begin{equation}
\log\left(\frac{\pi^\prime_{1}}{\pi^\prime_{2}}\right)^{\left(t+1\right)}=\log\left(\frac{\pi^\prime_{1}}{\pi^\prime_{2}}\right)^{\left(t\right)}+a_{t}\left(\delta_{1}-\delta_{2}\right)\left(\log\left(g\left(\theta\right)\right)-\log q_{\lambda}^{\left(K+1\right)}\left(\theta\right)\right),\label{eq:update pi}
\end{equation}
where $q^{(K+1)}_{\lambda}(\theta)$ is given in Eq. \eqref{eq:thevariationalmixturesapproximation} and the natural-gradient update for the new means $\mu_{K+1}$ is
\begin{equation}
\mu_{K+1}^{\left(t+1\right)}=\mu_{K+1}^{\left(t\right)}+a_{t}\delta_{2}\left(\beta_{K+1}^{\left(t\right)}\beta_{K+1}^{\left(t\right)^{\top}}+D_{K+1}^{2\left(t\right)}\right)\left(\nabla_{\theta}\log\left(g\left(\theta\right)\right)-\nabla_{\theta}\log q^{(K+1)}_{\lambda}\left(\theta\right)\right);\label{eq:update mu}
\end{equation}
 \begin{align*}
\delta_{1} &=\left(\sum_{k=1}^{K}\pi_{k}N\left(\varphi|\mu_{k},
\beta_{k}\beta_{k}^{\top}+D_{k}^{2}\right)\right)/\delta_{tot},\;\;\;
\delta_{2} =\left(N\left(\varphi|\mu_{K+1},\beta_{K+1}
\beta_{K+1}^{\top}+D_{K+1}^{2}\right)\right)/\delta_{tot},
\intertext{where}
\delta_{tot} &
=\left(\left(1-\pi_{K+1}\right)
\left(\sum_{k=1}^{K}\pi_{k}N\left(\varphi|\mu_{k},
\beta_{k}\beta_{k}^{\top}+D_{k}^{2}\right)\right)+\pi_{K+1}N
\left(\varphi|\mu_{K+1},\beta_{K+1}\beta_{K+1}^{\top}+D_{K+1}^{2}\right)\right).
\end{align*}
Updating the variational parameters $\beta_{K+1}$
and $d_{K+1}$ is now discussed. There are two reasons why the reparameterisation trick is not used to update the variational parameters $\beta_{K+1}$
and $d_{K+1}$. The first is that \citet{Miller2016} find that using the
 reparameterisation trick in the boosting variational method still results in a large variance and
it is necessary to use many samples to estimate the gradient of the variational lower bound. The second is that it may be impossible to use the reparameterisation trick because of the copula transformation. An alternative method to update the variational parameters $\beta_{K+1}$
and $d_{K+1}$ is now discussed.
%

The gradients of the lower bound in Eq. \eqref{eq:standardgradientlowerbound}
require the gradient $\nabla_{\lambda}\log q_{\lambda}\left(\theta\right)$.
The gradient of $\log q_{\lambda}\left(\theta\right)$
with respect to $\beta_{K+1}$ is
\begin{multline*}
\nabla_{\textrm{vech}\left(\beta_{K+1}\right)}\log q_{\lambda}\left(\theta\right)=\frac{\pi_{K+1}N\left(\varphi|\mu_{K+1},
\beta_{K+1}\beta_{K+1}^{\top}+D_{K+1}^{2}\right)}{\delta_{tot}}\\
\textrm{vech}\left(-\left(\beta_{K+1}\beta_{K+1}^{\top}+D_{K+1}^{2}\right)^{-1}
\beta_{K+1}+\right.\\
\left.\left(\beta_{K+1}\beta_{K+1}^{\top}+D_{K+1}^{2}\right)^{-1}
\left(\varphi-\mu_{K+1}\right)\left(\varphi-\mu_{K+1}\right)^{\top}
\left(\beta_{K+1}\beta_{K+1}^{\top}+D_{K+1}^{2}\right)^{-1}\beta_{K+1}\right),
\end{multline*}
and the gradient of $\log q_{\lambda}\left(\theta\right)$ with respect
to $d_{K+1}$ is
\begin{multline*}
\nabla_{d_{K+1}}\log q_{\lambda}\left(\theta\right)=
\frac{\pi_{K+1}N\left(\varphi|\mu_{K+1},\beta_{K+1}
\beta_{K+1}^{\top}+D_{K+1}^{2}\right)}{\delta_{tot}}\\
\textrm{diag}\left(-\left(\beta_{K+1}\beta_{K+1}^{\top}+
D_{K+1}^{2}\right)^{-1}D_{K+1}+\right.\\
\left.\left(\beta_{K+1}\beta_{K+1}^{\top}+D_{K+1}^{2}\right)^{-1}
\left(\varphi-\mu_{K+1}\right)\left(\varphi-\mu_{K+1}\right)^{\top}
\left(\beta_{K+1}\beta_{K+1}^{\top}+D_{K+1}^{2}\right)^{-1}D_{K+1}\right).
\end{multline*}
We also employ control variates as in \citet{Ranganath:2014} to reduce
the variance of an unbiased estimate of gradient of the $\nabla_{\textrm{vech}\left(\beta_{K+1}\right)}\mathcal{L}\left(\lambda\right)$
and $\nabla_{d_{K+1}}\mathcal{L}\left(\lambda\right)$ and the efficient
natural-gradient updates using the conjugate gradient methods given in
\citet{Tran2020}. To use a conjugate gradient linear solver to compute
$F_{\lambda}^{-1}\nabla_{\lambda}\mathcal{L}\left(\lambda\right)$
it is only necessary to be able to quickly compute matrix vector products of the
form $F_{\lambda}x$ for any vector $x$,
 without needing to
store the elements of $F_{\lambda}$. When $\beta_{K+1}$
is a vector, the natural gradient can be computed efficiently as outlined
in algorithm \ref{alg:Computing-the-natural gradient for beta and d} in section \ref{additionalalgorithm} of the online supplement. Our updates for $\beta_{K+1}$ and $d_{K+1}$ are pre-conditioned gradient steps based on the  natural gradient update for a Gaussian approximation in the added component, not the natural gradient in the mixture approximation.

\begin{algorithm}[H]
\caption{Variational Algorithm \label{alg:Variational-Algorithm}}

\begin{enumerate}
\item (a) Initialize $\lambda_{K+1}^{\left(0\right)}=\left(\mu_{K+1}^{\top \left(0\right)},
    \textrm{vech}(\beta_{K+1}^{\left(0\right)})^{\top},d_{K+1}^{\top \left(0\right)},
    \pi_{K+1}^{\left(0\right)}\right)$,
set $t=0$, and generate $\theta_{s}^{\left(t\right)}\sim q_{\lambda}^{\left(K+1\right)}\left(\theta\right)$ for $s=1,...,S$. Let $m_{\beta}$, $m_{d}$,
be the number of elements in $\textrm{vech}\left(\beta_{K+1}\right)$, and
$d_{K+1}$.

(b) Evaluate the control variates $\varsigma_{\textrm{vech}\left(\beta_{K+1}\right)}^{\left(t\right)}=\left(\varsigma_{1,\textrm{vech}\left(\beta_{K+1}\right)}^{\left(t\right)},...,\varsigma_{m_{\beta},\textrm{vech}\left(\beta_{K+1}\right)}^{\left(t\right)}\right)^{'}$,
$\varsigma_{d_{K+1}}^{\left(t\right)}=
\left(\varsigma_{1,d_{K+1}}^{\left(t\right)},...,
\varsigma_{m_{d},d_{K+1}}^{\left(t\right)}\right)^{'}$,
with
\begin{equation}
\varsigma_{i,d_{K+1}}^{\left(t\right)}=\frac{\wh {\textrm{Cov}}\left(\left[\log\left(\pi\left(\theta\right)\right)-\log q_{\lambda}\left(\theta\right)\right]\nabla_{\lambda_{i,d_{K+1}}}\log q_{\lambda}\left(\theta\right),\nabla_{\lambda_{i,d_{K+1}}}\log q_{\lambda}\left(\theta\right)\right)}{\wh\V\left(\nabla_{\lambda_{i,d_{K+1}}}\log q_{\lambda}\left(\theta\right)\right)},\label{eq:cv}
\end{equation}
for $i=1,...,m_{d}$, where $\wh {\textrm{Cov}}$ and $\wh\V\left(\cdot\right)$ are the
sample estimates of covariance and variance based on $S$ samples
from step (1a). The $\varsigma_{\textrm{vech}\left(\beta_{K+1}\right)}^{\left(t\right)}$
are estimated similarly.
\end{enumerate}
Repeat until the stopping rule is satisfied
\begin{itemize}
\item Update $\beta_{K+1}$, $d_{K+1}$:
\end{itemize}
\begin{enumerate}
\item Generate $\theta_{s}^{\left(t\right)}\sim q_{\lambda}^{\left(K+1\right)}\left(\theta\right)$ for $s=1,...,S$.
\item Compute $\widehat{\nabla_{\textrm{vech}\left(\beta_{K+1}\right)}\mathcal{L}\left(\lambda^{\left(t\right)}\right)}=\left(g_{1,\textrm{vech}\left(\beta_{K+1}\right)}^{\left(t\right)},...,g_{m_{\beta},\textrm{vech}\left(\beta_{K+1}\right)}^{\left(t\right)}\right)$,
with
\begin{equation}
g_{i,\textrm{vech}\left(\beta_{K+1}\right)}^{\left(t\right)}=\frac{1}{S}\sum_{s=1}^{S}\left[\log\left(\pi\left(\theta_{s}^{\left(t\right)}\right)\right)-\log q_{\lambda}\left(\theta_{s}^{\left(t\right)}\right)-\varsigma_{i,\textrm{vech}\left(\beta_{K+1}\right)}^{\left(t-1\right)}\right]\nabla_{\lambda_{i,\textrm{vech}\left(\beta_{K+1}\right)}}\log q_{\lambda}\left(\theta_{s}\right)\label{eq:gradLB_beta}
\end{equation}

\item The gradient of lower bound $\widehat{\nabla_{d_{K+1}}\mathcal{L}\left(\lambda^{\left(t\right)}\right)}=\left(g_{1,d_{K+1}}^{\left(t\right)},...,g_{m_{d},d_{K+1}}^{\left(t\right)}\right)$
can be computed similarly as in Eq. \eqref{eq:gradLB_beta}.
\item Compute the control variate $\varsigma_{\textrm{vech}\left(\beta_{K+1}\right)}^{\left(t\right)}$
and $\varsigma_{d_{K+1}}^{\left(t\right)}$ as in Eq. \eqref{eq:cv}
and compute $g_{\beta_{K+1}}^{\textrm{nat}}$ and $g_{d_{K+1}}^{\textrm{nat}}$ using
algorithm \ref{alg:Computing-the-natural gradient for beta and d} in section  \ref{additionalalgorithm} of the online supplement.
\item Compute $\triangle d_{K+1}$, $\triangle \textrm{vech}\left(\beta_{K+1}\right)$
using ADAM as described in section \ref{subsec:Learning-Rate} of the online supplement. Then, set
$d_{K+1}^{\left(t+1\right)}=d_{K+1}^{\left(t\right)}+\triangle d_{K+1}$,
$\textrm{vech}\left(\beta_{K+1}\right)^{\left(t+1\right)}=\textrm{vech}\left(\beta_{K+1}\right)^{\left(t\right)}+\triangle \textrm{vech}\left(\beta_{K+1}\right)$.
\end{enumerate}
\begin{itemize}
\item Update $\mu_{K+1}$ and $\pi_{K+1}$
\end{itemize}
\begin{enumerate}
\item Generate $\theta_{s}^{\left(t\right)}\sim q_{\lambda}^{\left(K+1\right)}\left(\theta\right)$ for $s=1,...,S$.
\item Use Eq. \eqref{eq:update pi} to update $\pi_{K+1}$ and Eq. \eqref{eq:update mu}
to update $\mu_{K+1}$, respectively. Set $t=t+1$
\end{enumerate}
\end{algorithm}

However, we find this pre-conditioned gradient step improves efficiency compared to the ordinary gradient. 
Section~\ref{efficiencyNaturalGradients} gives further details.
Algorithm \ref{alg:Variational-Algorithm} presents the full variational algorithm.

\subsection{Initialising a New Mixture Component\label{subsec:Initialising-a-New}}

Introducing a new component requires  setting the initial values
for the new component parameters $\left(\mu_{K+1},\beta_{K+1},D_{K+1}\right)$
and the new mixing weight $\pi_{K+1}$. A good initial value for the
new mixture component should be located in the region of the target
posterior distribution $\pi\left(\theta\right)$ that is not well represented
by the existing mixture approximation $q_{\lambda}^{\left(K\right)}\left(\theta\right)$.
There are many ways to set the initial value. We now discuss some 
that work well in our examples. The elements in $\beta_{K+1}$ are
initialized randomly by drawing from $N\left(0,0.001^{2}\right)$, the diagonal elements
in $D_{K+1}$ are initialized by 0.001. These values ensure that the optimisation algorithm is stable because generated values will be close to $\mu_{K+1}$ in the first few initial iterations.
The mixture weight $\pi_{K+1}$
is initialized by 0.5. Algorithm \ref{alg:initial values for mu_K+1}
gives the initial value for  $\mu_{K+1}=\left(\mu_{1,K+1},...,\mu_{m,K+1}\right)^{\top}$.


\begin{algorithm} [H]
\caption{Initial values for $\mu_{K+1}$. \label{alg:initial values for mu_K+1}}

Input: $\left\{ \pi_{k},\mu_{k},\beta_{k},d_{k}\right\} _{k=1}^{K}$
and $\gamma$

Output: initial values for $\mu_{K+1}$
\begin{itemize}

\item For $i=1$ to $m$
\begin{itemize}
\item Construct a grid of $S$ values of the $i$th transformed parameter $\varphi_{i}$. One way to construct the
grid of $S$ values is to draw samples from the current approximation,
and compute the minimum and maximum values $\left(\min\left(\varphi_{i}\right),\max\left(\varphi_{i}\right)\right)$
for $\varphi_{i}$. Let $\varphi_{s}^{*}$ be a vector containing $\varphi_{i,s}$ and other transformed parameters fixed at their means or some other reasonable values. 
\item Compute $\theta_{j,s}^{*}=t_{\gamma}^{-1}\left(\varphi_{j,s}^{*}\right)$
for $s=1,...,S$ and $j=1,...,m$.
\item Compute the $w_{s}^{*}=\frac{g\left(\theta_{s}^{*}\right)}
    {q_{\lambda}^{\left(K\right)}\left(\theta_{s}^{*}\right)}$
for $s=1,...,S$.
\item Set
$\mu_{i,K+1}=\varphi_{i,s}$ with a probability proportional to
the weight $w_{s}^{*}$.
\end{itemize}
\item Alternatively, when the dimension of the parameters is large,
\begin{itemize}
\item Draw $S$ samples from the current variational approximation $q_{\lambda}^{\left(K\right)}\left(\theta\right)$,
and compute  the weights $w_{s}^{*}=
\pi\left(\theta_{s}^{*}\right)/q_{\lambda}^{\left(K\right)}\left(\theta_{s}^{*}\right)$
for $s=1,...,S$. Then, set $\mu_{K+1}=\varphi_{s}^{*}$ with a probability proportional
to the weight $w_{s}^{*}$.
\end{itemize}
\end{itemize}
\end{algorithm}


%



\section{Examples\label{sec:Examples}}

To illustrate the performance of the proposed variational approximations described in section \ref{sec:Variational-Approximation}, we employ them to approximate
complex and high-dimensional distributions, where their greater flexibility may
increase the accuracy of inference and prediction compared to
simpler approximations.



The section has four examples. The first approximates a high dimensional, skewed, and heavy tailed distribution. The second example approximates a high dimensional multimodal distribution. The third example approximates the posterior distributions of the model parameters of a logistic regression model with a complex prior distribution. The fourth example fits a Bayesian deep neural network regression model. In all the examples, the following variational approximations are considered:
\begin{itemize}
\item (A1) A mixture of Gaussians variational approximation (MGVA). The Gaussian variational approximation is a special case with $K=1$.
\item (A2) The copula-based mixture of Gaussians variational approximation (CMGVA). The Gaussian copula variational approximation is a special case where $K=1$.
\item (A3) The mean-field mixture of Gaussians variational approximation (MGVA-MF). We use the terms mean-field variational approximation to refer to the case where the covariance matrix for each component in the mixture is diagonal.
\item (A4) The mean-field copula-based mixture of Gaussians variational approximation (CMGVA-MF).
\item (A5) The mixture of skew Gaussian variational approximation (MSGVA). This is a special case of the variational approximation described in Eq. \eqref{eq:skewnormalva} when the variational parameters $\gamma_{i}=1$ for all $i=1,...,m$. 
\item (A6) The mixture of skew Gaussian copula-based variational approximation (CMSGVA) given in Eq. \eqref{eq:skewnormalva}. For MSGVA and CMSGVA, only the first component is the skew Gaussian distribution. The other $K-1$ components are Gaussian distributions. 
\end{itemize}
All the examples
are implemented in Matlab. The first three examples were run on a standard desktop computer. The fourth example was run on a 28 CPU-cores of a high performance computer cluster.
Unless otherwise stated, we use the estimates of the variational lower
bound to select the best variational approximations. In principle,
making the variational family more flexible by adding new components should not  reduce
the variational lower bound; in practice, the difficulty of the optimization means that adding a new component may worsen the approximation.
The variational approximation is useful when the exact MCMC method is impossible or computationally expensive.
The boosting approach, where  an existing approximation is improved by adding one new component at a time allows us to tune the accuracy/computational effort trade-off, where we start with a rough fast approximation and keep improving until the computational budget is exhausted.  
All the variational parameters are initialised using the approach
in section \ref{subsec:Initialising-a-New}. 



\subsection{Skewed and Heavy-Tailed High-Dimensional Distributions \label{subsec:Multivariate t-distribution}}

This section investigates the ability of variational approximations (A1)-(A6) to approximate skewed, heavy-tailed, and  high-dimensional target distributions. The true target distributions $\pi\left(\theta\right)$ are assumed to follow multivariate t-copula,
\begin{equation}
\pi\left(\theta\right):=\pi\left(\zeta\right)\prod_{i=1}^{m}\bigg |\frac{d\zeta_{i}}{d\theta_{i}}\bigg |.
\end{equation}
The density $\pi\left(\zeta\right)$ is a
multivariate $t$-distribution with zero mean, full-covariance matrix
(ones on the diagonal and $0.8$ on the off-diagonals), and degrees
of freedom $df=4$. The Yeo-Johnson (YJ) transformation with parameters set to 0.5 \citep{Yeo2000}
is used. The dimension of the parameters
$\theta$ is set to $m=100$.
The number of factors $r_{1}$ is set to $4$ for the first component
and $r_{k}=1$, for each additional mixture component for $k=2,...,20$ for MGVA, CMGVA, MSGVA, and CMSGVA.
We use $S=100$ samples to estimate the lower bound and the gradients of the lower bound.
The algorithm  in \citet{Smith2020} is performed for $5000$ iterations
to obtain the optimal variational parameters for the first component
of the mixture,  and then  algorithm \ref{alg:Variational-Algorithm}
is performed for 5000 iterations to obtain the optimal variational
parameters for each additional component of the mixture for variational approximations (A1)-(A6).

\begin{figure}[H]
\caption{Plot of the average lower bound values over the last 500 steps for variational approximations (A1)-(A6) for the 100-dimensional
multivariate $t$-copula example.\label{CopMixNom}}

\centering{}\includegraphics[width=15cm,height=8cm]{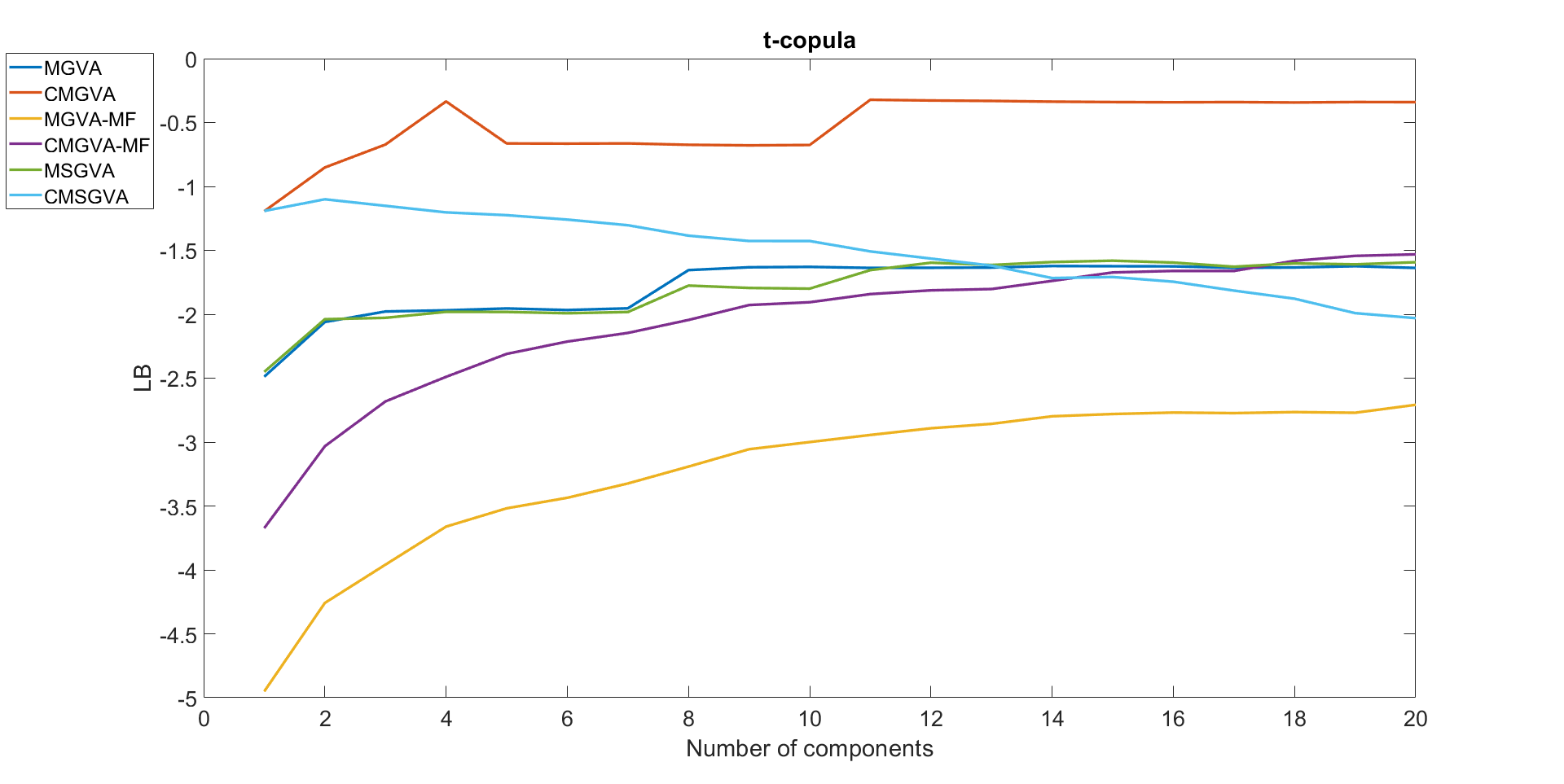}
\end{figure}

Figure \ref{CopMixNom}
shows the average lower bound value over the last 500 steps of the
optimisation algorithm for variational approximations (A1)-(A6). 
The figure shows that the Gaussian and skew Gaussian copula-based estimators have similar lower bounds that are larger than the Gaussian, skew Gaussian, mean-field Gaussian copula-based approximations, and the mean-field Gaussian variational approximation. As expected, the mean-field Gaussian variational approximation has the lowest lower bound value because it does not capture the dependence structure of the $\theta$ posterior. The figure also shows
that there is substantial improvement obtained by going from $k=2$ to $8$ components for most variational approximations, and there are no significant improvements thereafter. Interestingly, the lower bound values of the CMSGVA decrease when more components are added in the mixture. The Gaussian copula has greater lower bound than MGVA for any number of components for this example. The CMGVA with $k=4$ components has a significantly higher lower bound compared to other variational approximations for this example. 

\begin{figure}[H]
\caption{Left: Kernel density estimates of the marginal parameter $\theta_{1}$ approximated by the mean-field Gaussian, Gaussian, skew Gaussian, Gaussian copula, mean-field Gaussian copula, skew Gaussian copula variational approximations. Right: Kernel density estimates of the marginal parameter $\theta_{1}$ approximated by the Gaussian, Gaussian copula, 3-component CMGVA, 4-component CMGVA, and 8-component MGVA \label{2_20CMGVA_texample}}

\centering{}\includegraphics[width=15cm,height=8cm]{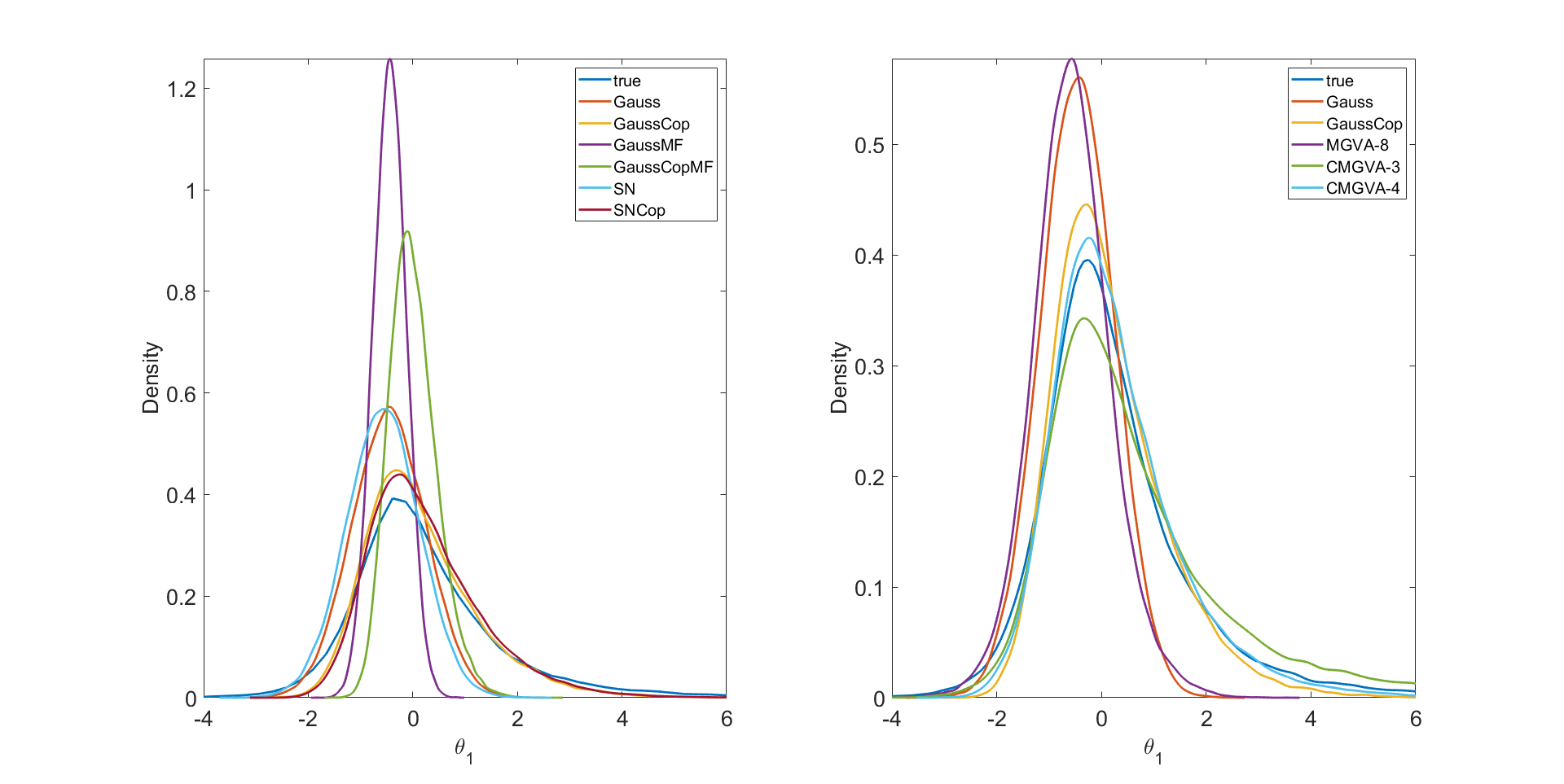}
\end{figure}

Figure \ref{2_20CMGVA_texample}
shows the kernel density estimates of the marginal densities of the
parameter $\theta_{1}$ estimated using the different variational approximations.
The left panel of 
figure~\ref{2_20CMGVA_texample} compares the performance of mean-field Gaussian, mean-field Gaussian copula, Gaussian, Gaussian copula, skew Gaussian, and skew Gaussian copula variational approximations. It shows that the mean-field Gaussian and mean-field Gaussian copula variational approximations significantly underestimate the posterior variances of the parameter $\theta_{1}$. The Gaussian copula and skew Gaussian copula perform better than the Gaussian and skew Gaussian variational approximations. The right panel of figure \ref{2_20CMGVA_texample} shows that the CMGVA with $k=3$ and $k=4$ components captures both the skewness and the heavy tails of the marginal posteriors $\pi\left(\theta_{1}\right)$ better than the Gaussian copula  and the mixture of the Gaussians variational approximations with $k=8$ components. 

\begin{figure}[H]
\caption{Left: Scatter plot of the observations from the first and second margins generated from true target density (red), Gaussian copula variational approximation (yellow), and 3-component CMGVA (blue). Right:  Scatter plot of the observations from the first and second margins generated from true target density (blue), Gaussian copula variational approximation (yellow), and 4-component CMGVA (red). \label{2_20CMGVA_texample_scatter}}

\centering{}\includegraphics[width=15cm,height=8cm]{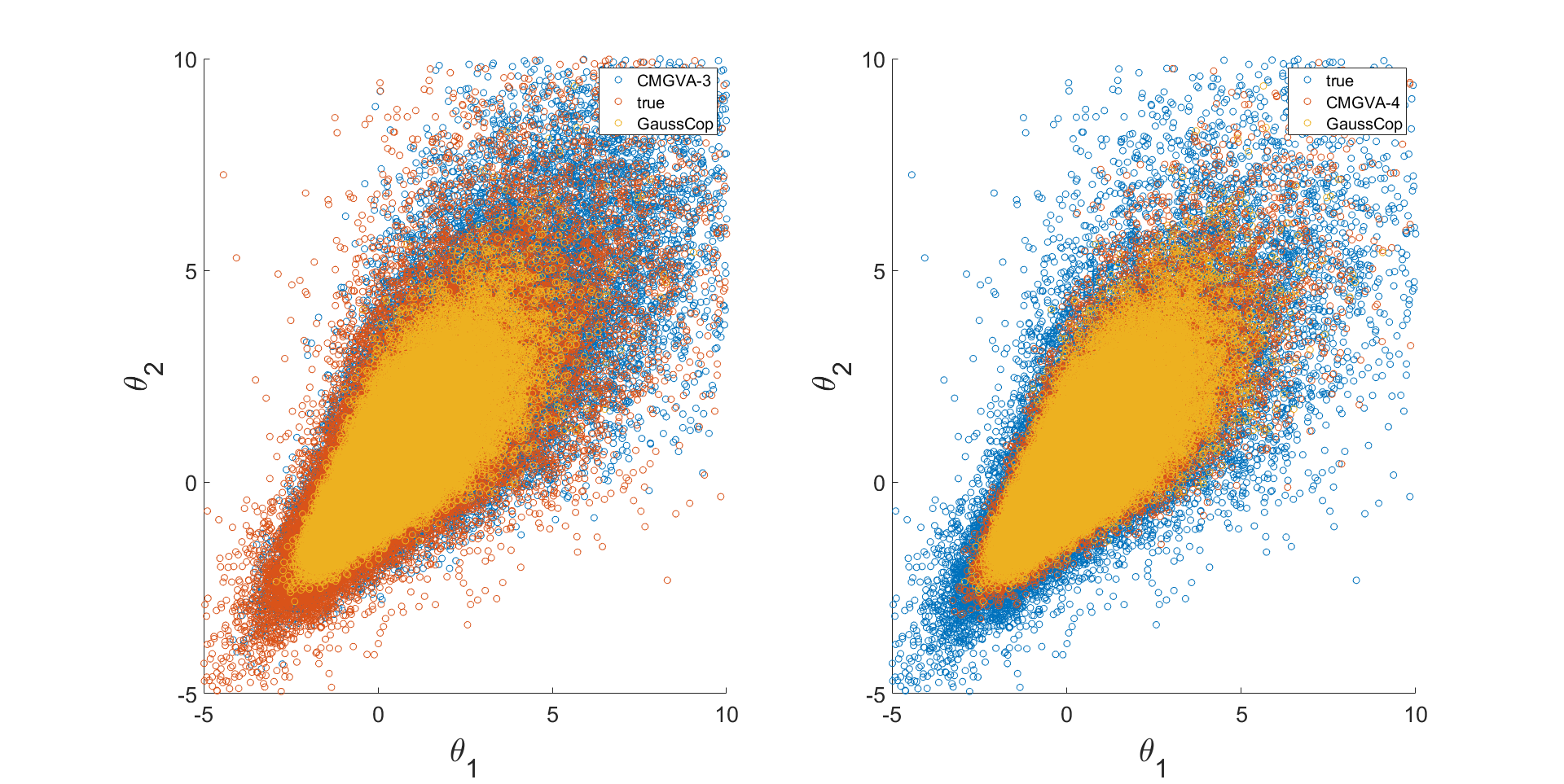}
\end{figure}

Figure~\ref{2_20CMGVA_texample_scatter}
shows the scatter plot of the observations from the first and second
margins generated from the true target densities, Gaussian copula variational
approximation, the $3$-component CMGVA, and the $4$-component CMGVA. This suggests that the CMGVA is better at capturing the skewness and heavy-tailed properties of the true target densities
compared to the Gaussian copula-based variational approximations.

We now compare the accuracy of the CMGVA to the densities estimated using the Hamiltonian Monte Carlo (HMC) method of \citet{hoffman2014no}, called the No U-Turn Sampler (NUTS); this method is a popular MCMC algorithm for sampling high dimensional posterior distributions. For all examples,
the NUTS tuning parameters, such as the number of leapfrog steps and the step size, are set to the default values as in the STAN reference manual \footnote{$https://mc-stan.org$}. We ran the HMC method  for $1005000$ iterations, discarding the initial $5000$ iterations as warm-up. The remaining $1000000$ MCMC samples are stored for further analysis. The inefficiency of the HMC method is measured using the integrated autocorrelation time (IACT) defined in section \ref{IACT_section} of the online supplement.

\begin{figure}[H]
\caption{Kernel Density Estimates of the marginal parameter $\theta_{1}$ estimated using the HMC method, 3-component CMGVA, and 4-component CMGVA for $m=100$. \label{2_20CMGVA_texample_MCMC}}
\centering{}\includegraphics[width=15cm,height=8cm]{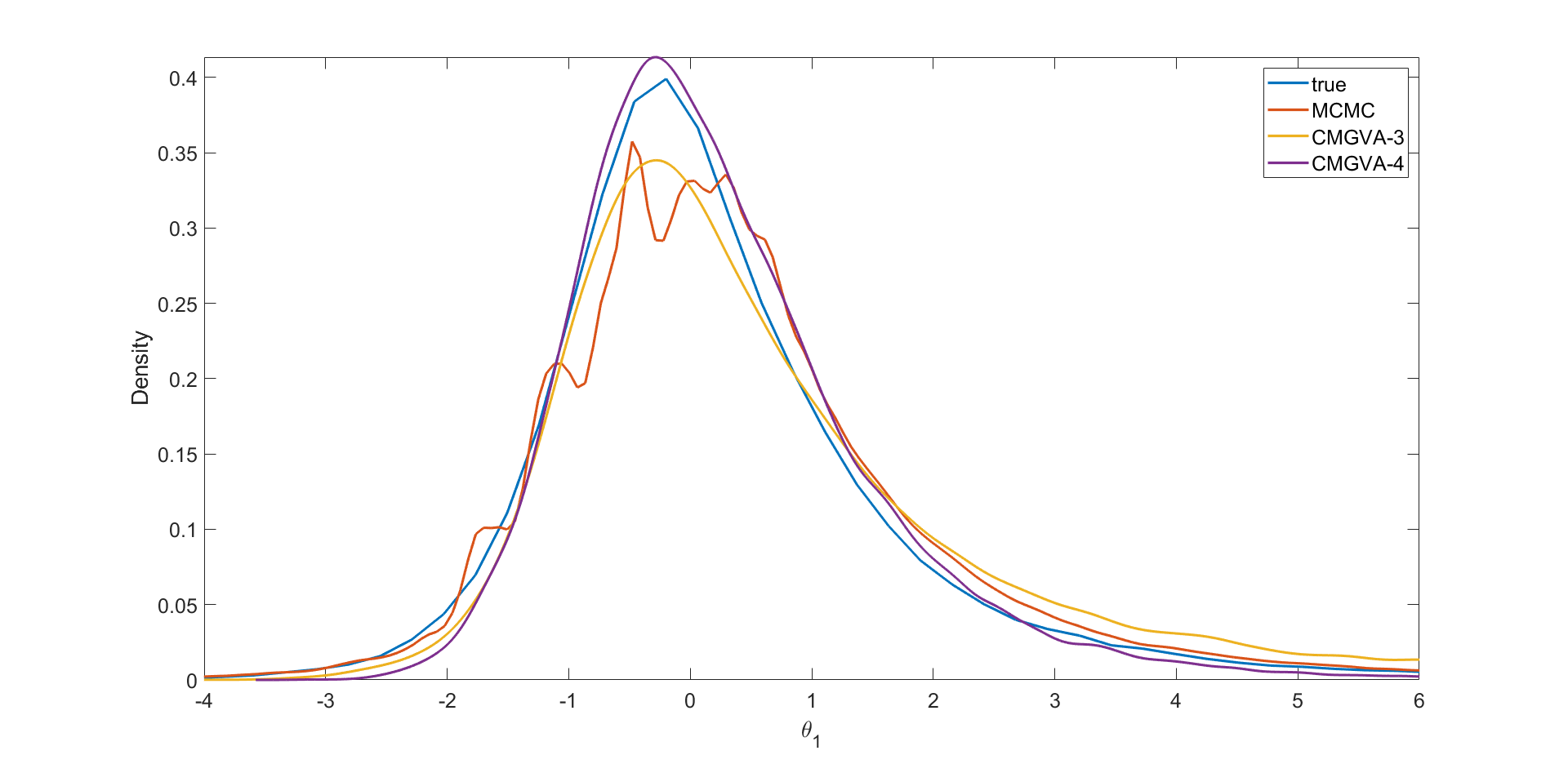}
\end{figure}

Figure \ref{fig:MCMC_TDIST_HMC_BOXPLOT} in section \ref{additionalfigureskewed} of the online supplement shows the IACT of the parameters $\theta$ estimated using the HMC method. The figure shows that the average IACT of the parameters is $391$ with the average effective sample size of $2556.70$. Figure \ref{fig:MCMC_TDIST_HMC_TRACE} also shows that the Markov chain sometimes gets stuck, indicating that the HMC method performs quite inefficiently in this example. Figure \ref{2_20CMGVA_texample_MCMC} shows that the 3-component and 4-component CMGVA are more accurate than the HMC method for estimating the marginal density of the parameter $\theta_1$. The same applies to other marginal densities. Again, this suggests that the CMGVA can capture the skewness and heavy-tailed properties of the true target densities.

The CPU time for the HMC method is 167.5 minutes. The time taken for estimating the 3- and 4-component CMGVA are 36.15 minutes and 50.10 minutes\footnote{The CPU time for estimating the 3-component CMGVA is the total time taken for estimating 1 to 3-component CMGVA. Similar calculations are used for calculating the CPU time for the 4-component CMGVA.}. Therefore, the total CPU time for estimating the 3- and 4-component CMGVA are 4.63 and 3.34 times faster than the HMC method.




%



\subsection{Multimodal High-Dimensional Distributions \label{subsec:Multivariate-Mixture-of normals}}
This section investigates the ability of the proposed variational approximations (A1)--(A6) to approximate multimodal high-dimensional target distributions.
The true target distribution is the
multivariate mixture of normals, $\pi\left(\theta\right)=\sum_{c=1}^{3}w_{c}N\left(\theta|u_{c},\Sigma_{c}\right).$
The dimension of the parameters $\theta$ is set to $100$. Each element $u_{i,c}$ is uniformly drawn from the interval $\left[-2,2\right]$
for $i=1,...,100$ and $c=1,..,3$. We set the full covariance matrix
$\Sigma_{c}$ with ones on the diagonal and the correlation coefficients
$\rho=0.2$ and $0.8$ in the off-diagonals for all components.
The number of factors $r_{1}=4$ for the first component
and $r_{k}=1$, for each additional mixture component for $k=2,...,20$ for variational approximations A1, A2, A5, and A6.
Similarly to the previous example, we use $S=100$ samples to estimate the lower bound values and the gradients
of the lower bound. The algorithm in \citet{Smith2020} is performed
for $5000$ iterations to obtain the optimal variational parameters
for the first component of the mixture and then algorithm~\ref{alg:Variational-Algorithm}
is performed for $5000$ iterations to obtain the optimal variational
parameters for each additional component of the mixture. The step sizes are set to the values given in section \ref{subsec:Learning-Rate} of the online supplement.


Figure \ref{CopMixNomMixNomExample} in section \ref{additionalfigure} of the online supplement shows the average lower bound values over the last 500 steps for the variational approximations (A1)-(A6) for the 100-dimensional mixture of normals example. The figure shows that the performances of the CMGVA, MGVA, MSGVA, and CMSGVA are comparable for the cases $\rho=0.2$ and $\rho=0.8$. Figure \ref{YJ_plot_mixnom} in section \ref{additionalfigure} of the online supplement confirms that by showing that all the YJ-parameters are close to 1 for CMGVA and CMSGVA and all the $\widetilde{\alpha}$ parameters in Eq. \eqref{eq:skewnormalva}  are close to 0 for MSGVA and CMSGVA for the case $\rho=0.8$. Similar conclusions hold for the case $\rho=0.2$. The mixture of normals is a special case of CMGVA when all the
YJ-parameters are equal to 1, are special cases of MSGVA when all the $\widetilde{\alpha}$ parameters are close to 0, and are special cases of CMSGVA when all the YJ-parameters are close to 1 and all the $\widetilde{\alpha}$ are close to 0. The MGVA, CMGVA, MSGVA, and CMSGVA are clearly much better than MGVA-MF and CMGVA-MF in this example.

\begin{figure}[H]
\caption{Kernel density estimates of some of the marginal parameters $\theta$ approximated
with Gaussian, Gaussian Copula, skew Gaussian, skew Gaussian copula, and the optimal variational approximations (A1)-(A6) together with the true marginal distributions and the marginal distributions
estimated using the HMC method
for the mixture of normals example with $\rho=0.8$ \label{fig:Kernel-Density-Estimates mixnormal mix100dim08}}

\centering{}\includegraphics[width=15cm,height=5cm]{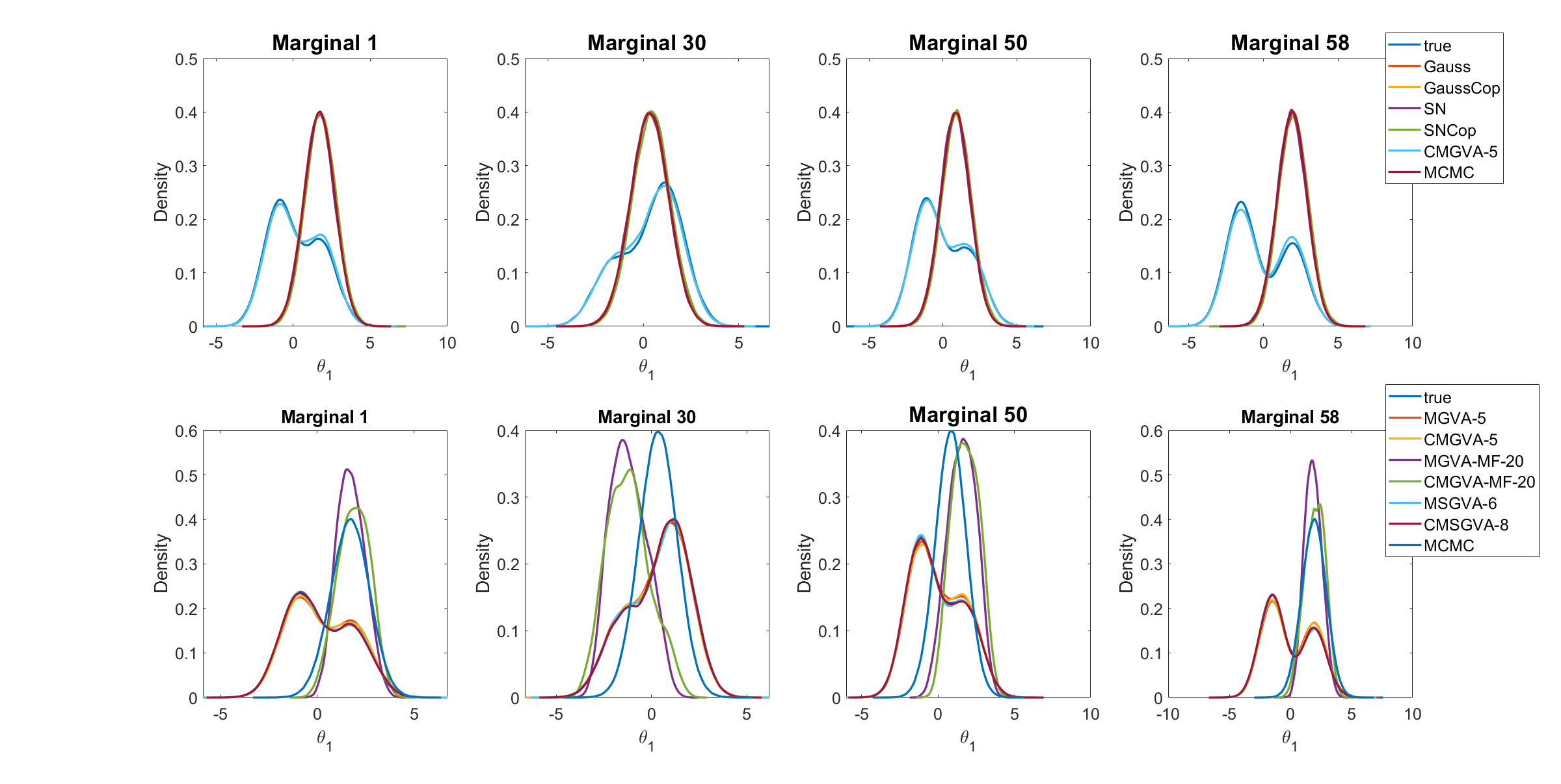}
\end{figure}

The top panel of figure \ref{fig:Kernel-Density-Estimates mixnormal mix100dim08}
shows the kernel density estimates of some of the posterior densities  
of the marginal parameters of $\theta$
approximated with several of the variational approximations,
together with the true marginal distributions and the marginal distributions estimated using the HMC method of \citet{hoffman2014no} for the case $\rho=0.8$. The HMC method ran for $1005000$ iterations, 
with the initial $5000$ iterations discarded as warm up. 
The remaining $1000000$ iterations are used for further analysis.
Figure~\ref{fig:MCMC_MIXNOM_HMC_BOXPLOT} in section~\ref{additionalfigure} of
the online supplement shows the IACT of the parameters $\theta$ estimated using HMC. The figure shows that the average IACT of the parameters is $226.32$ with the average effective sample size of $4418.52$, indicating
that the HMC method is also quite inefficient for this example.

The bottom panel of
figure~\ref{fig:Kernel-Density-Estimates mixnormal mix100dim08} shows the
variational approximations (A1)-(A6), together with the true marginal
distributions and the marginal distributions estimated using HMC for
$\rho=0.8$. 
The top panel shows that the Gaussian, Gaussian copula, skew Gaussian, skew Gaussian copula, and HMC approaches are unable to approximate multimodal distributions. The bottom panel shows that the optimal MGVA, CMGVA, MSGVA, and CMSGVA perform much better than the optimal MGVA-MF and CMGVA-MF.   
Similar conclusions can be made from 
figure~\ref{fig:Kernel-Density-Estimates mixnormal mix100dim02} in 
section~\ref{additionalfigure} of the online supplement for the case $\rho=0.2$. The CPU time for the HMC method for this example is $83.75$ minutes. The time taken to estimate the 5-component CMGVA is 37.25 minutes which is $2$ and a quarter times faster than HMC.

\begin{figure}[H]
\caption{Top: Scatter plot of the observations generated from true target distribution
(blue), and a 5-component CMGVA (orange) for the mixture of normals
example with $\rho=0.8$; Bottom: Scatter plot of the observations
generated from true target distribution (blue), and the Gaussian copula (orange)
for the mixture of normals example with $\rho=0.8$ \label{fig:bivariatescatterplotmixnormal08}}

\centering{}\includegraphics[width=15cm,height=8cm]{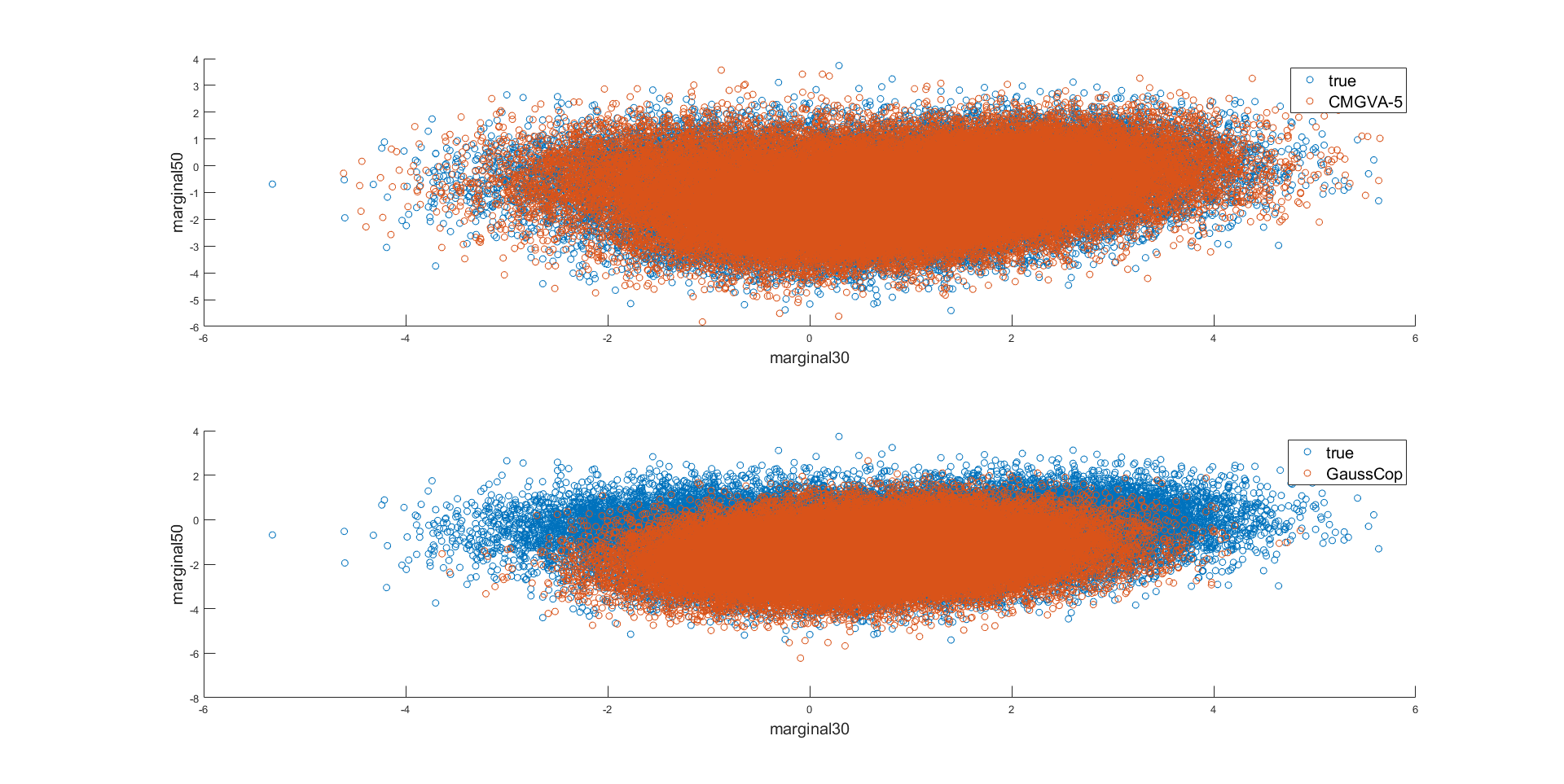}
\end{figure}

Finally, figures~\ref{fig:bivariatescatterplotmixnormal08} and 
\ref{fig:bivariatescatterplotmixnormal02} in section~\ref{additionalfigure} 
of the online supplement show the scatter plots
of the observations generated from
the true density, the Gaussian copula variational approximation, and
the optimal CMGVA for $\rho=0.8$ and $0.2$, respectively. The
figures confirm that the CMGVA can capture the bimodality and complex-shaped
of the two dimensional distribution of the parameters.

We now show that the CMGVA does not overfit a Gaussian target distribution, which we take as a $q$-variate normal
with zero mean and full covariance matrix with ones on the diagonal and correlation coefficients $\rho=0.8$ in the off-diagonals.
Figure~\ref{CopNomExample} in section~\ref{additionalfigure} of the online supplement plots the average lower bound values over the last 500 steps for the CMGVA for this example and shows that no improvement is obtained by adding additional components in the mixture.

The two examples in sections \ref{subsec:Multivariate t-distribution} and \ref{subsec:Multivariate-Mixture-of normals} suggest that: (1)~The CMGVA can approximate heavy tails, multimodality, skewness and other
complex properties of the high dimensional target distributions, outperforming
the Gaussian copula and other variational approximations.
Section~\ref{subsec:Multivariate t-distribution} shows that the Gaussian copula outperforms MGVA, MGVA-MF, CMGVA-MF, MSGVA, and CMSGVA at approximating a skewed target distribution. 
The optimal variational approximations (A1)-(A6) are better than a Gaussian copula at approximating multimodal target distributions.
(2)~Adding a few components to the MGVA, CMGVA, MSGVA, and CMSGVA generally improves their ability to approximate
complex target distributions. Therefore, the proposed approach
can be considered as a refinement of the Gaussian copula and skew Gaussian copula
variational approximations. (3) Adding additional components one at a time provides
a practical  method for constructing  an increasingly
complicated  approximation and applies to a variety
of multivariate target distributions. (4) The HMC method of \citet{hoffman2014no} fails to estimate the high dimensional, heavy tailed, and multimodal target distribution.

%

\subsection{Bayesian Logistic Regression Models with Complex Prior Distributions \label{subsec:RegressionModelWithComplexPriorDistributions}}

This section considers a logistic regression model 
\[
p\left(y_{i}|x_{i},b\right)=\frac{\exp\left(y_{i}x_{i}^{\top}b\right)}{1+\exp\left(y_{i}x_{i}^{\top}b\right)}, 
\]
with the response $y_{i}\in\left\{ 0,1\right\} $, and 
with a complex prior distribution for the regression parameters $b=(b_0,b_1,...,b_p)^{\top}$. 
The prior for each regression parameter, except the intercept, is
the two-component mixture of skew normals
\begin{equation}
p\left(b_{i}|w,\sigma_{1}^{2},\sigma_{2}^{2},\alpha\right)=wSN\left(b_{i};0,\sigma_{1}^{2},\alpha\right)+\left(1-w\right)SN\left(b_{i};0,\sigma_{2}^{2},\alpha\right),\;i=1,...,p;
\end{equation}
$SN\left(b_{i};0,\sigma^{2},\alpha\right)$ is the skew-normal distribution
of \citet{azzalini85} with density $\frac{2}{\sigma}\phi\left(\frac{b_{i}}{\sigma}\right)\Phi\left(\alpha\frac{b_{i}}{\sigma}\right)$.
We set $w=0.5$, $\sigma_{1}^{2}=0.01$, $\sigma_{2}^{2}=100$, and
$\alpha=-4$. This prior is motivated by a variable selection scenario,
where some coefficients may be 0 and we would like to set these close
to zero. The prior for the intercept term $b_0$ is $N(0,1)$



We consider the spam, krkp, ionosphere, and mushroom data for the logistic regression model; they
have sample
sizes $n=4601$, $351$, $3196$, and $8124$, with $104$, $111$,
$37$, and $95$ covariates, respectively and are also considered by \citet{Ong2018} and \citet{Smith2020};
the data are available
from the UCI Machine Learning Repository \citep{Lichman2013}\footnote{see https://archive.ics.uci.edu/ml/datasets.php for further details.}. 
In the results reported below, we include all covariates but only use the first $50$ observations of each dataset.
The small dataset size and the complex prior distribution for each regression parameter are chosen to create a complex posterior structure to evaluate the performance of variational approximations (A1)-(A6) when the posteriors are non-Gaussian. 

Similarly to the previous example, we set the number of factors $r_{1}$ to $4$ for the first component
and $r_{k}=1$, for each additional mixture component for $k=2,...,20$ for variational approximations A1, A2, A5, and A6.
We use $S=100$ samples to estimate the lower bound values and the gradients
of the lower bound. The algorithm in~\citet{Smith2020} is performed
for 5000 iterations to obtain the optimal variational parameters
for the first component of the mixture and then algorithm \ref{alg:Variational-Algorithm}
is performed for 5000 iterations to obtain the optimal variational
parameters for each additional component of the mixture.

\begin{figure}
\caption{Plots of the average lower bound values over the last 500 steps
for the variational approximations (A1)-(A6) for
the Bayesian logistic regression model for the four datasets \label{fig:LB_Logistic_regression_ionosphere_krkp}}

\centering{}\includegraphics[width=15cm,height=8cm]{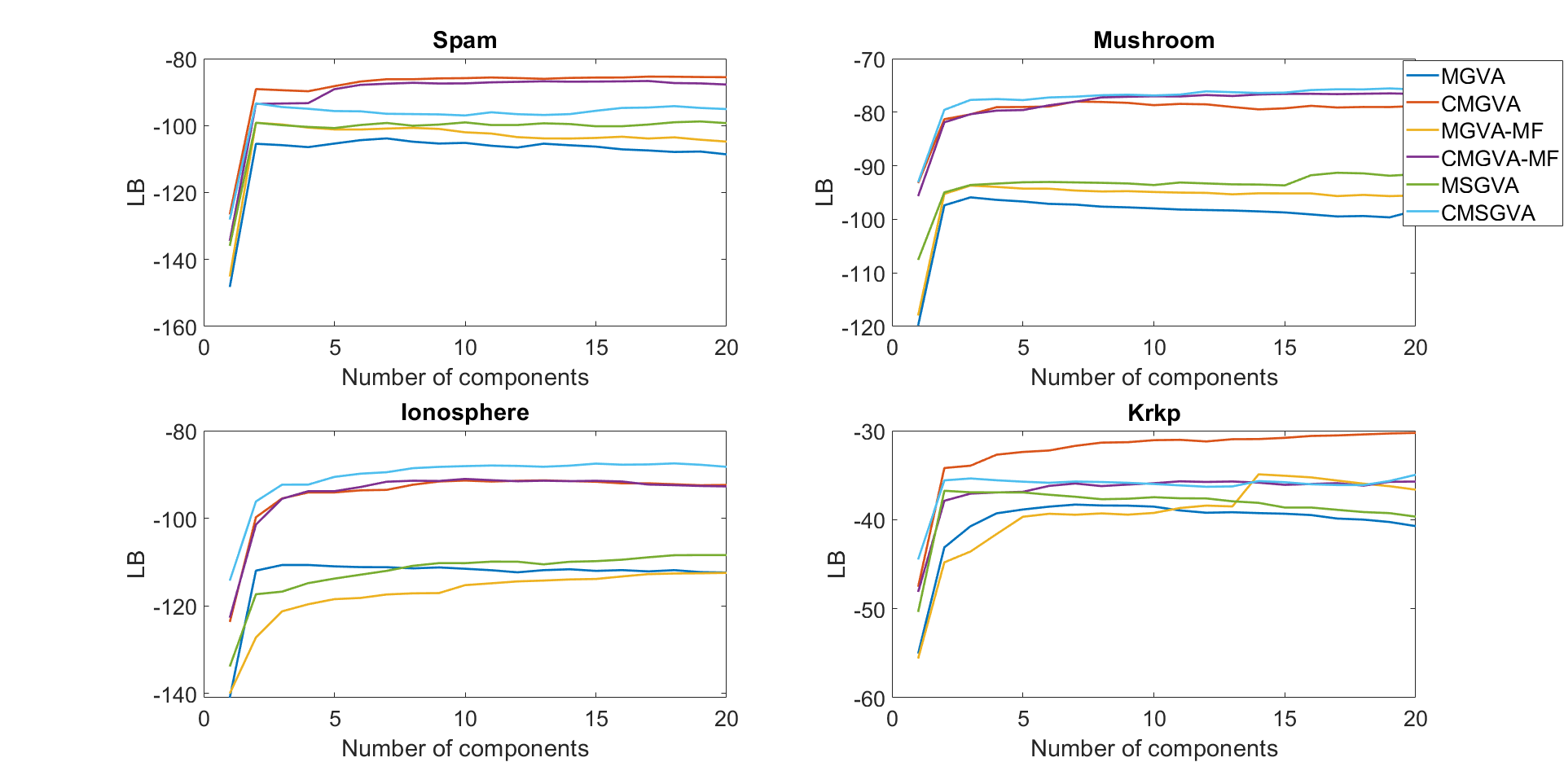}
\end{figure}

Figure~\ref{fig:LB_Logistic_regression_ionosphere_krkp} shows the average lower bound values over the last 500 steps of the
optimisation algorithm for the variational
approximations (A1)-(A6) for the Bayesian logistic regression model for the four
datasets. The Gaussian and skew Gaussian copula variational approximations outperform the Gaussian and skew Gaussian variational approximations.  Interestingly, the mean-field Gaussian copula variational approximation is better than the skew Gaussian and the Gaussian variational approximation for the spam, mushroom, and ionosphere datasets. 
The figure also shows that adding a few components in the mixture for variational approximations (A1)-(A6) improves the lower bound values for all datasets. The optimal CMGVA performs the best for the spam and krkp datasets. The CMSGVA performs slightly better than CMGVA for the mushroom and ionosphere datasets.  



%
%
%

\subsection{Flexible Bayesian Regression with a Deep Neural Network \label{subsec:Bayesian-Deep-Neural}}

Deep feedforward neural network (DFNN) models
with binary and continuous response variables are widely used
for classification and regression in the machine learning literature.
The DFNN method can be viewed as a way to efficiently transform a vector
of $p$ raw covariates $X=\left(X_{1},...,X_{p}\right)^{\top}$ into
a new vector $Z$ having the  form
\begin{equation}
Z := f_{L}\left(W_{L,}f_{L-1}\left(W_{L-1},...,f_{1}\left(W_{1},X\right)\right)\right).
\end{equation}
Each $Z_{l}=f_{l}\left(W_{l},Z_{l-1}\right)$, $l=1,...,L$, is called
a hidden layer, $L$ is the number of hidden layers in the network,
$W=\left(W_{1},...,W_{L}\right)$ is the set of weights and $Z_{0}=X$
by construction. The function $f_{l}\left(W_{l},Z_{l-1}\right)$ is
assumed to be of the form $h_{l}\left(W_{l}Z_{l-1}\right)$, where
$W_{l}$ is a matrix of weights that connect layer $l-1$ to layer $l$, which includes weight coefficients attached to the input $Z_{l-1}$ and the constant terms,
and $h_{l}\left(\cdot\right)$ is a scalar activation function. Estimation
in complex high dimensional models like DFNN regression models is
challenging. This section studies the accuracy of posterior densities and the
predictive performance of the variational approximations (A1)-(A6) for a DFNN regression model with continuous responses; see \citet[chapters 6, 9, and 10]{Goodfellow2016} for a
comprehensive recent discussion of DFNNs and other types of neural networks.



Consider a dataset $D$ with $n$ observations,  with $y_{i}$ the scalar
response and $x_{i}=\left(x_{i1},...,x_{ip}\right)^{\top}$ the vector
of $p$ covariates. We consider a neural network structure with the
input vector $x$ and a scalar output. Denote $z_{l}:=f_{l}\left(x,w\right)$,
$l=1,...,M$, the units in the last hidden layer, $w$ is the vector of weights
up to the last hidden layer, and $b=\left(b_{0},b_{1},...,b_{M}\right)^{\top}$
are the weights that connect the variable $z_{l}$, $l=1,...,M$,
to the output $y$. The model, with a continuous response $y$, can be written as
\begin{eqnarray}
\tau^{2} & \sim & Gamma\left(1,10\right),\\
w_{i} & \sim & 0.5SN\left(0,\sigma_{1}^{2},\alpha\right)+0.5SN\left(0,\sigma_{2}^{2},\alpha\right),\;\textrm{for}\;i=1,...,M_{w},\\
b_{l} & \sim & 0.5SN\left(0,\sigma_{1}^{2},\alpha\right)+0.5SN\left(0,\sigma_{2}^{2},\alpha\right),\;\textrm{for}\;l=1,...,M,\\
y|x,w,b,\tau & \sim & N\left(b_0+\widetilde{b}^{\top}z,1/\tau^{2}\right),
\end{eqnarray}
where $z=(z_1,...,z_M)^{\top}$, $\widetilde{b}=(b_1,...,b_M)^{\top}$, $M_w$ is the number of weight parameters; $SN\left(B;0,\sigma^{2},\alpha\right)$ is a skew-normal density of \citet{azzalini85} defined in Section \ref{subsec:RegressionModelWithComplexPriorDistributions}, and $Gamma\left(1,10\right)$ is the gamma distribution 
with shape parameter 1 and scale parameter 10. We set  $\sigma_{1}^{2}=0.01$, $\sigma_{2}^{2}=100$, and
$\alpha=-4$. \citet{Miller2016} uses similar priors for $\tau^2$. The priors for $b_{l}$ for $l=1,...,M$ and $w_i$ for $i=1,...,M_w$ will shrink some of the coefficients that may be 0 or very close to zero.


%

All the examples use the rectified linear unit (ReLU) $h\left(x\right):=\max\left(0,x\right)$ as an activation function, 
unless otherwise stated; ReLU is widely applied in the deep learning literature 
\citep{Goodfellow2016} because it is easy to use within optimization as it is  quite similar to a linear function,
except that it  outputs zero for negative values of $x$.

We consider the auto and abalone datasets.
The auto dataset, available from \citet{James2013}, consists of 392 observations
for different makes of cars, with the response being gas mileage in miles per 
gallon, with 7 additional covariates used here to predict the mileage. 
The abalone dataset,
available on the UCI Machine Learning Repository, has 4177 observations.
The response variable is the number of rings used to determine the
age of the abalone. There are 9 covariates including sex, length, diameter,
as well as other measurements of the abalone.  We use 90\% of the data for training and the rest for computing the log of the approximate predictive score in Eq.~\eqref{eq:estimatepredictivedensity}. Both datasets have continuous responses


Neural nets with (8,5,5,1), (8,10,10,1) and (8,20,20,1) structures are used for the auto dataset.
For the (8,5,5,1) structure, the input
layer has 8 variables, there are two hidden layers each having 5 units and there is a scalar output. The first layer has $8 \times 5 = 40$ $w$ parameters, the second layer has
$6 \times 5 = 30$ $w$ parameters (including the intercept term), and $6$ $b$ parameters (including the intercept term); this gives a total of $76$ parameters. Similar calculations can be made for the (8,10,10,1) and (8,20,20,1) structures to give a total of $201$ and $601$ parameters, respectively.
Neural nets with (9,5,5,1), (9,10,10,1) and (9,20,20,1) structures are used for the abalone data set. Similar calculations can be done for them to give a total of $75$, $211$, and $621$ parameters, respectively.

This section studies the accuracy of the inference and the 
predictive performance for variational approximations A1 and A2. For this example, variational approximations A5 and A6 are not implemented due to the numerical issues in estimating the skew normal and skew normal copula variational approximations. Sections~\ref{subsec:Multivariate t-distribution} to \ref{subsec:RegressionModelWithComplexPriorDistributions} show that CMGVA is better than MSGVA and CMSGVA. We do not compare the posterior density of the parameters obtained from the variational approximations A1 and A2 to HMC, as it is difficult to obtain the exact posterior distribution for the parameters of the Bayesian neural network. \citet{Papamarkou2021} shows that the Markov chains generated by the Metropolis-Hastings and Hamiltonian Monte Carlo methods fail to converge for estimating the parameters of the Bayesian neural network model. To show the lack of convergence of the HMC method, we ran it for $1005000$ iterations discarding the initial $5000$ iterations as warm up for estimating neural nets with a (9,10,10,1) structure for the auto dataset. 
Figure~\ref{fig:MCMC_NN_HMC_BOXPLOT} in section~
\ref{additionalfiguresNN} of the online supplement shows the IACT of the parameters $\theta$ estimated using the HMC method. The figure shows that the average IACT of the parameters is $1904.08$ with an average effective sample size of $525.19$, indicating that HMC is very inefficient for this example.
In addition, figure~\ref{fig:MCMC_NN_ABALONE} in 
section~\ref{additionalfiguresNN} of the online supplement shows the trace plots of the parameters of the neural net with the (9,10,10,1) structure for the auto dataset. The figure shows that the parameters mix poorly.  
Computing time for a 10-component CMGVA is an order of magnitude less than for the HMC.





To evaluate the predictive accuracy of a DFNN regression model estimated
by the variational approximations A1 and A2, we consider
the posterior predictive density defined as
\begin{equation}
p\left(y|x,D\right)=\int p\left(y|x,\theta\right)p\left(\theta|D\right)d\theta.
\end{equation}
Given that we have the variational approximation of the posterior
distribution, we can define the approximate predictive density 
\begin{equation}
g\left(y|x,D\right)=\int p\left(y|x,\theta\right)q_{\lambda}\left(\theta\right)d\theta.\label{eq:approximatepredictive}
\end{equation}
Computing the approximate posterior predictive density in Eq. (\ref{eq:approximatepredictive})
is challenging because it involves high dimensional integrals that
cannot be solved analytically. However, it can be estimated using
Monte Carlo integration. The estimate of the log of the approximate posterior
predictive score is 
\begin{equation}
\textrm{log}\, \widehat{g}\left(y|x,D\right)=\textrm{log}\left(\frac{1}{R}\sum_{r=1}^{R}p\left(y|x,\theta^{r}\right)\right),\;\theta^{r}\sim q_{\lambda}\left(\theta\right). \label{eq:estimatepredictivedensity}
\end{equation}
The higher the log of the approximate posterior predictive score, the more  accurate the prediction. 

In this example, the number of factors is set to 1 for all $20$
components in the mixture; $S=200$ samples are used to estimate
the gradients of the lower bound; $R=10000$ samples are used to estimate the log of the approximate posterior predictive score in 
Eq.~\eqref{eq:estimatepredictivedensity}; section~\ref{stoppingNN} of the online supplement discusses the stopping criterion for the optimisation algorithm.

\begin{figure}
\caption{Top panels: The plots of the average lower bound values over the last 100 steps for the variational approximations A1 and A2 for the (8,10,10,1) neural net structure for the auto dataset and the (9,10,10,1) neural net structure for the abalone dataset. Bottom panels: The plots of the log of the estimated approximate posterior predictive scores for the variational approximations A1 and A2 for the (8,10,10,1) neural net structure for the auto dataset and the (9,10,10,1) neural net structure for the abalone dataset. \label{fig:LB_pred_1010}}

\centering{}\includegraphics[width=15cm,height=8cm]{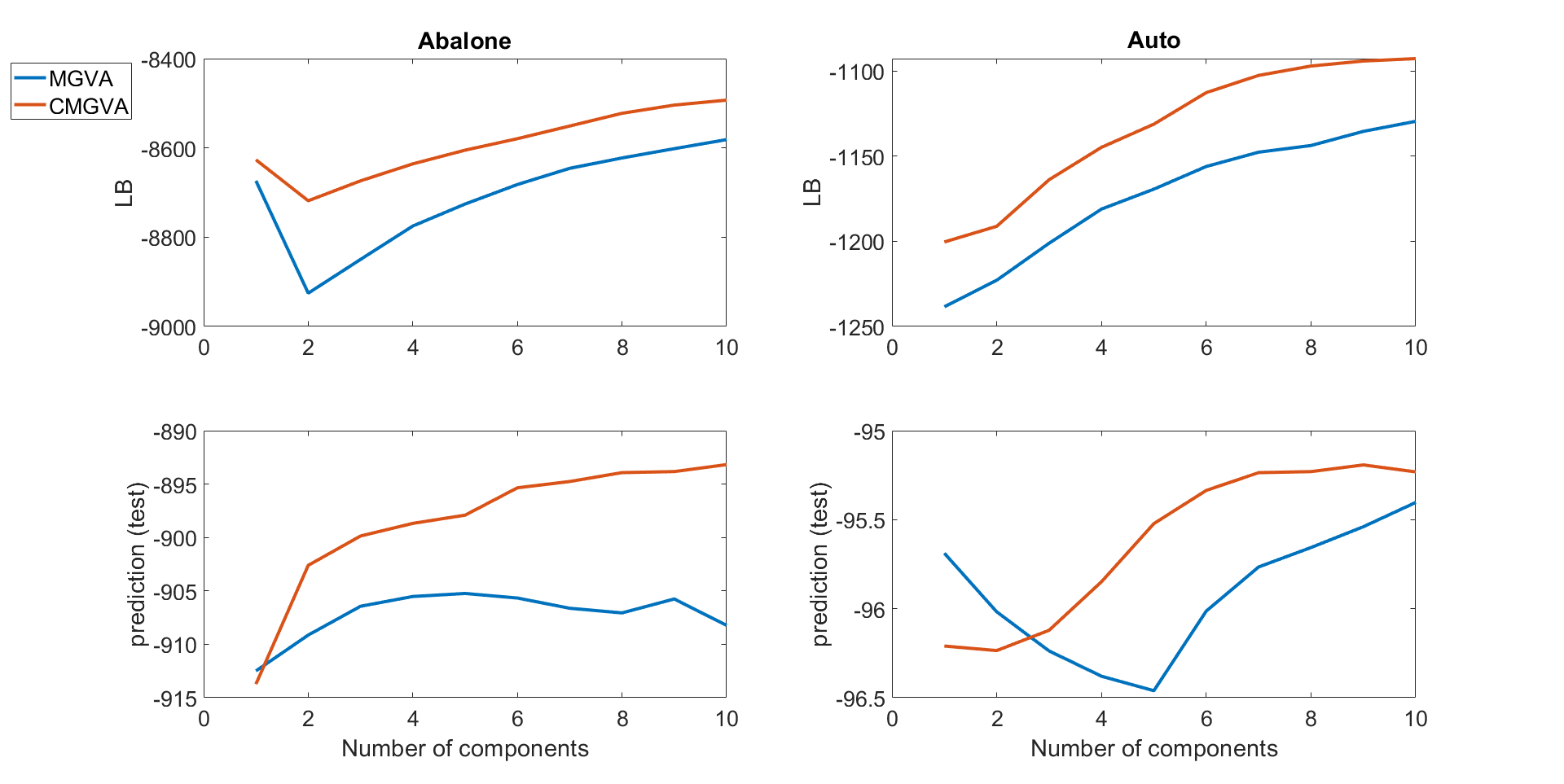}
\end{figure}

\begin{figure}
\caption{Top panels: The plots of the average lower bound values over the last 100 steps for the variational approximations A1 and A2 for the (8,20,20,1) neural net structure for the auto dataset and the (9,20,20,1) neural net structure for the abalone dataset. Bottom panels: The plots of the log of the estimated approximate posterior predictive scores for the variational approximations A1 and A2 for the (8,20,20,1) neural net structure for the auto dataset and the (9,20,20,1) neural net structure for the abalone dataset. \label{fig:LB_pred_2020}}

\centering{}\includegraphics[width=15cm,height=8cm]{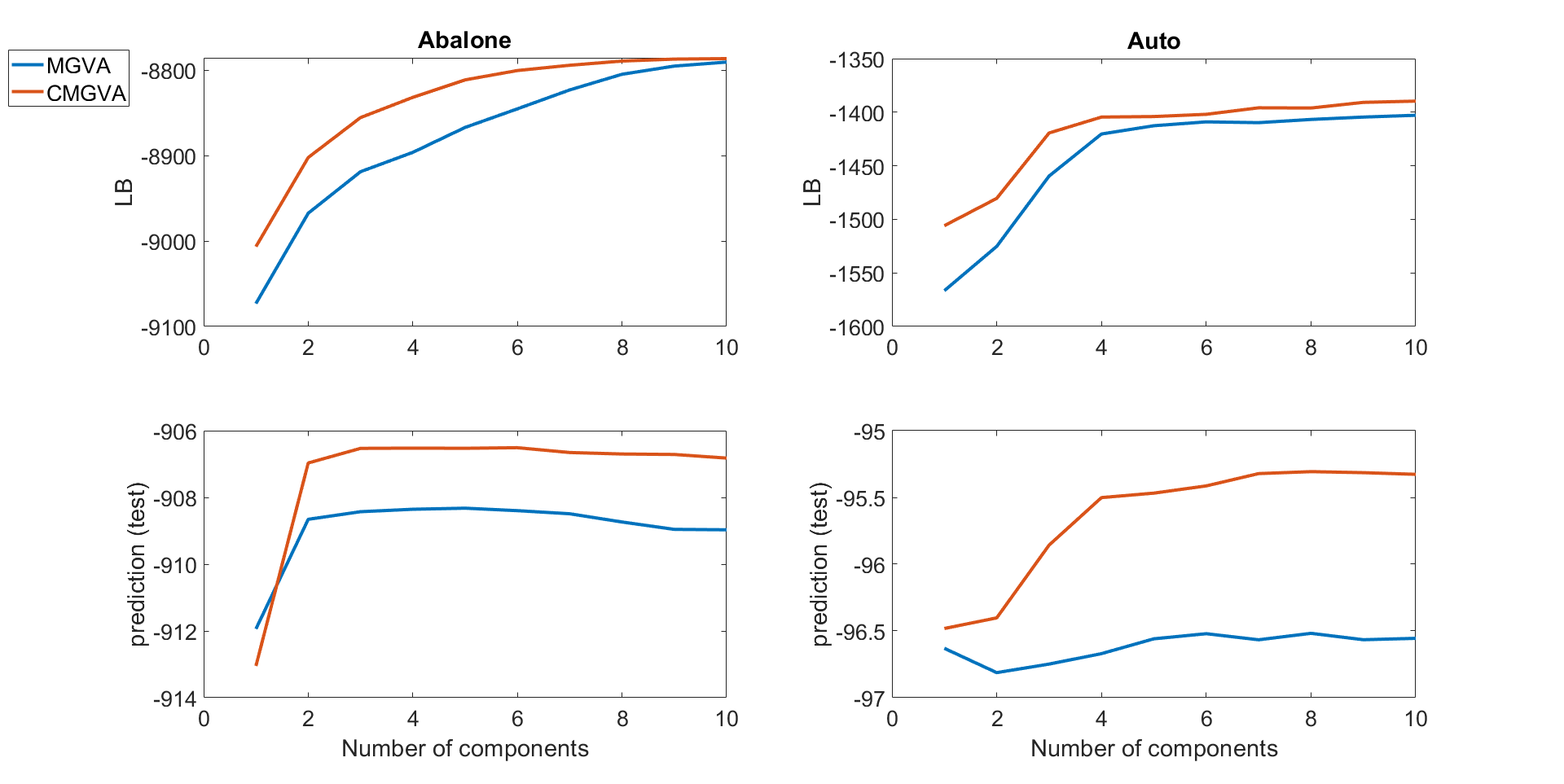}
\end{figure}

The top panels of figure~\ref{fig:LB_pred_55} in 
section~\ref{additionalfiguresNN} of the online supplement show the average
lower bound values over the last 100 steps of the optimisation algorithm for the (8,5,5,1) neural net structure for the auto dataset and the (9,5,5,1) structure for the abalone dataset. The figure shows that adding components to the variational approximations A1 and A2 increases the lower
bound significantly for both datasets. Clearly, CMGVA performs best for both datasets. The lower panels of figure \ref{fig:LB_pred_55} show the log of the estimated approximate posterior predictive scores evaluated for the test data. The figure also shows that adding more components to the variational approximations can improve the prediction accuracy significantly. Similar conclusions can be drawn from figure \ref{fig:LB_pred_1010} for the (8,10,10,1) neural net structure for the auto dataset and the (9,10,10,1) neural net structure for the abalone dataset and figure \ref{fig:LB_pred_2020} for the (8,20,20,1) neural net structure for the auto dataset and the (9,20,20,1) neural net structure for the abalone dataset. This suggests the usefulness of the proposed variational approximations for complex and high-dimensional Bayesian deep neural network regression models. 


We now compare the optimal CMGVA to the planar flows of \citet{rezende2015variational} with flow lengths of $10$ transformations. Tanh is the non-linearity function with the initial distribution being Gaussian  with mean $\mu$ and a diagonal covariance matrix; $S=1000$ samples are used to accurately estimate
the gradients of the lower bound. A similar stopping rule to that described in
section~\ref{stoppingNN} of the online supplement is used. 
Figures~\ref{fig:The-plots-ofnn5} and \ref{fig:The-plots-ofnn2020} in 
section~\ref{planarflows} of the online supplement show that the lower 
bound of the planar flows for the two
datasets for neural nets with different structures increase at the
start and then converge. Table~\ref{CMGVAagainstPlanar} in 
section~\ref{additionalfiguresNN} of the online supplement shows that the CMGVA has a higher lower bound and higher log of the approximate posterior predictive scores compared to the planar flows.


\subsection{Efficiency of the Natural Gradient \label{efficiencyNaturalGradients}}
\begin{figure}[H]
\caption{The plot of the lower bound values over iterations for the ordinary 
and natural gradient methods for the 2-components CMGVA for the 100-dimensional
multivariate t-copula, multivariate mixture of normals, Bayesian logistic regression model (spam data), and the DFNN regression model for the Abalone dataset with a (9,5,5,1) neural net structure examples. \label{fig:ordinarynaturalgradients}}
\centering{}\includegraphics[width=15cm,height=8cm]{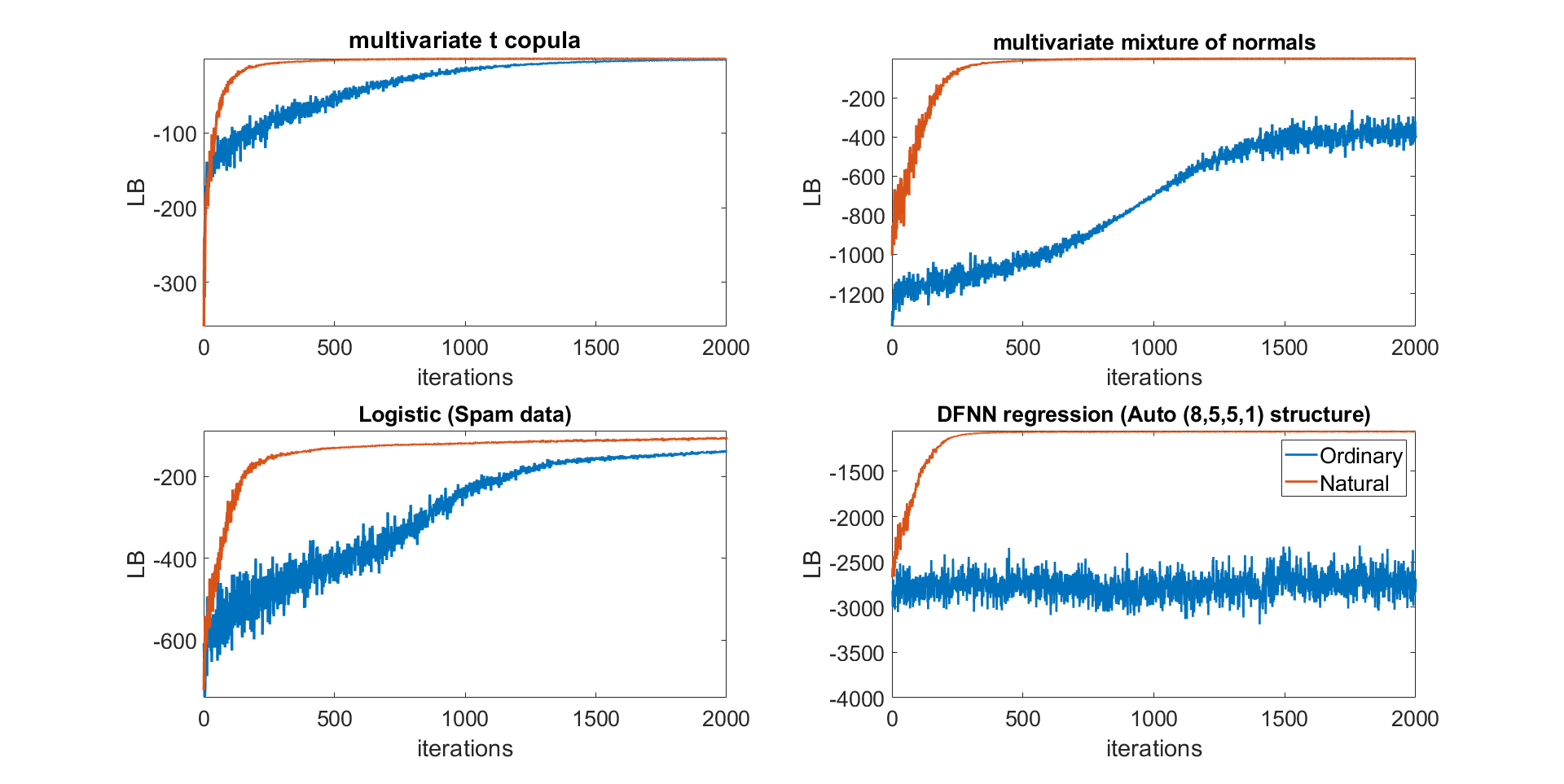}
\end{figure}

This section compares the performance of the natural gradient and the ordinary gradient methods using the same initial values for the variational parameters for both methods. Figure~\ref{fig:ordinarynaturalgradients} shows the lower bound values over iterations for both methods for the 2-component CMGVA for the multivariate $t$-copula, multivariate mixture of normals, logistic regression (spam dataset), and Bayesian  DFNN regression (abalone dataset with neural net (9, 5, 5, 1) structure). The figure shows that the natural gradient is much less noisy and converges much faster than the ordinary gradient.



\section{Conclusion \label{sec:Conclusions}}

The article proposes flexible variational approximations based on a copula of a mixture of normals and constructs the computational algorithms for estimating the approximation. 
An important part of the approach is the construction of appropriate transformations of the parameters to try and simplify the joint posterior which is then estimated by a mixture of normals. 

The VB method is made efficient by using the natural gradient and control variates.
Our approach of adding one component at a time provides
a practical variational inference method that constructs an increasingly
complicated posterior approximation and is an extension and refinement
of state-of-the-art Gaussian and skew Gaussian copula variational approximations in
\citet{Smith2020}.  The proposed variational approximations apply to a wide range of Bayesian models;
we apply it to four
complex examples, including Bayesian deep learning regression models,
and show that it  improves upon the Gaussian copula and mixture of normals variational
approximations in terms of both inference and prediction. Our article uses a factor structure for the covariance matrix in our variational
approximation, but it is straightforward to extend the variational
approach to consider other sparse forms of the covariance structure, such
as a sparse Cholesky factorisation as in \citet{Tan2019}. 

We note that it is straightforward to use any other copula-based approximation as a first component in our approach, such as a $t$-copula. This may result in a simpler mixture  approximation than using the Gaussian copula as a first component.


\section{Acknowledgement}
The research of Robert Kohn was partially supported by an ARC Center of Excellence grant CE140100049.

\section{Online Supplement}
CopMixJCGSSupp.pdf (pdf file) provides additional examples, results and technical details. Computer code to implement the methods for the examples of this article is available online.

\clearpage

\renewcommand{\thealgorithm}{S\arabic{algorithm}}
\renewcommand{\theequation}{S\arabic{equation}}
\renewcommand{\thesection}{S\arabic{section}}
\renewcommand{\thepage}{S\arabic{page}}
\renewcommand{\thetable}{S\arabic{table}}
\renewcommand{\thefigure}{S\arabic{figure}}
\setcounter{page}{1}
\setcounter{section}{0}
\setcounter{equation}{0}
\setcounter{algorithm}{0}
\setcounter{table}{0}
\setcounter{figure}{0}
\def\E{\mathrm{E}}
\def\Var{\mathrm{V}}
\def\Cov{\mathrm{Cov}}
\def\({\Big ( }
\def\){\Big )}
\def\ACEMS{\textit{Australian Center of Excellence for Mathematical and Statistical Frontiers}}
\def\bs{\boldsymbol}

\setcounter{page}{0}

\title{\bf Online Supplement: Flexible Variational Bayes based on a Copula of a Mixture}
\maketitle
\section{Learning Rate \label{subsec:Learning-Rate}}

Setting the learning rate in a stochastic gradient algorithm is
very challenging, especially when the  parameter vector
is high dimensional. The choice of learning rate affects both the
rate of convergence and the quality of the optimum attained. Learning
rates that are too high can cause unstable optimisation,
while learning rates that are too low result in slow convergence
and can lead to a situation where the parameters erroneously
appear to have converged.
In all our examples, the learning rates are set adaptively using the
ADAM method \citep{Kingma2015} that gives different step sizes for
each element of the variational parameters $\lambda$. At iteration
$t+1$, the variational parameter $\lambda$ is updated as
\[
\lambda^{\left(t+1\right)}:=\lambda^{\left(t\right)}+\triangle^{\left(t\right)}.
\]
Let $g_{t}^{\textrm{nat}}$ denote the natural stochastic gradient estimate
at iteration $t$. ADAM computes (biased) first and second moment
estimates of the gradients using exponential moving averages,
\begin{eqnarray*}
m_{t} & = & \tau_{1}m_{t-1}+\left(1-\tau_{1}\right)g_{t}^{\textrm{nat}},\\
v_{t} & = & \tau_{2}v_{t-1}+\left(1-\tau_{2}\right)\left(g_{t}^{2}\right)^{\textrm{nat}},
\end{eqnarray*}
where $\tau_{1},\tau_{2}\in\left[0,1\right)$ control the decay rates.
The biased first and second moment estimates are corrected
by \citep{Kingma2015}
\begin{eqnarray*}
\widehat{m}_{t} & = & m_{t}/\left(1-\tau_{1}^{t}\right), \quad
\widehat{v}_{t}  =  v_{t}/\left(1-\tau_{2}^{t}\right);
\end{eqnarray*}
the change $\triangle^{\left(t\right)}$ is then computed as
\begin{equation*}
\triangle^{\left(t\right)}=\frac{\alpha\widehat{m}_{t}}{\sqrt{\widehat{v}_{t}}+\eps}.
\end{equation*}
We set $\tau_{1}=0.9$, $\tau_{2}=0.99$, and $\eps=10^{-8}$
\citep{Kingma2015}. It is possible to use different $\alpha$ for $\mu$, $\beta$, $d$, and $\pi$. We set $\alpha_{\mu}=0.01$, $\alpha_{\beta}=\alpha_{d}=\alpha_{\pi}=0.001$, unless stated otherwise.

\section{Algorithms\label{additionalalgorithm}}

Algorithm \ref{alg:Computing-the-natural gradient for beta and d} gives the formula for computing the natural gradients for the vectors $\beta_{K+1}$ and $d_{K+1}$.

\begin{algorithm}[H]
\caption{Computing the natural gradients for the vectors $\beta_{K+1}$ and
$d_{K+1}$. \label{alg:Computing-the-natural gradient for beta and d}}

Input: vectors $\beta_{K+1}$ and $d_{K+1}$ and the standard gradients
$g_{\beta_{K+1}}=\nabla_{\textrm{vech}\left(\beta_{K+1}\right)}\mathcal{L}\left(\lambda\right)$
and $g_{d_{K+1}}=\nabla_{d_{K+1}}\mathcal{L}\left(\lambda\right)$

Output: $g_{\beta_{K+1}}^{\textrm{nat}}=F_{\lambda}^{-1}g_{\beta_{K+1}}=\nabla_{\textrm{vech}\left(\beta_{K+1}\right)}^{\textrm{nat}}\mathcal{L}\left(\lambda\right)$
and $g_{d_{K+1}}^{\textrm{nat}}=F_{\lambda}^{-1}g_{d_{K+1}}=\nabla_{d_{K+1}}^{\textrm{nat}}\mathcal{L}\left(\lambda\right)$.
\begin{itemize}
\item Compute the vectors: $v_{1}=d_{K+1}^{2}-2\beta_{K+1}^{2}\circ d_{K+1}^{-4}$,
$v_{2}=\beta_{K+1}^{2}\circ d_{K+1}^{-3}$, and the scalars $\kappa_{1}=\sum_{i=1}^{m}\beta_{K+1}^{2}/d_{K+1}^{2}$,
and $\kappa_{2}=0.5\left(1+\sum_{i=1}^{m}v_{2i}^{2}/v_{1i}\right)^{-1}$.
\item Compute:
\begin{equation}
g_{\beta_{K+1}}^{\textrm{nat}}=\frac{1+\kappa_{1}}{2\kappa_{1}}\left(\left(g_{\beta_{K+1}}^{\top}\beta_{K+1}\right)\beta_{K+1}+d_{K+1}^{2}\circ g_{\beta_{K+1}}\right),
\end{equation}
 and
\begin{equation}
g_{d_{K+1}}^{\textrm{nat}}=0.5v_{1}^{-1}\circ g_{d_{K+1}}+\kappa_{2}\left[\left(v_{1}^{-1}\circ v_{2}\right)^{\top}g_{d_{K+1}}\right]\left(v_{1}^{-1}\circ v_{2}\right).
\end{equation}
\end{itemize}
\end{algorithm}

\section{Integrated Autocorrelation Time\label{IACT_section}}

To define our measure of the inefficiency of an MCMC sampler, we  define the integrated autocorrelation
time (IACT) for a univariate function $\psi(\theta)$ of parameter $\theta$ as
\[
\textrm{IACT}_{\psi}=1+2\sum_{j=1}^{\infty}\rho_{j,\psi},
\]
where $\rho_{j,\psi}$ is the $j$th autocorrelation of the iterates of $\psi(\theta)$
in the MCMC after the chain has converged. 
It measures the inefficiency of the sampling
scheme in terms of the multiple of its draws that are required to obtain the
same variance as an independent sampling scheme, e.g. if IACT = 10, then
we need ten times as many iterates as an independent scheme.
We use the CODA package of
\citet{Plummer2006} to estimate the IACT values of the parameters.
A low value of the IACT estimate suggests that the Markov chain mixes
well. 

\section{Additional Figures for the Skewed and Heavy-Tailed High-Dimensional Target Distribution Example \label{additionalfigureskewed}}

\begin{figure}[H]
\caption{The trace plots of some of the marginal distributions of the skewed and heavy-tailed high-dimensional target distribution in section \ref{subsec:Multivariate t-distribution}.
\label{fig:MCMC_TDIST_HMC_TRACE}}
\centering{}\includegraphics[width=15cm,height=8cm]{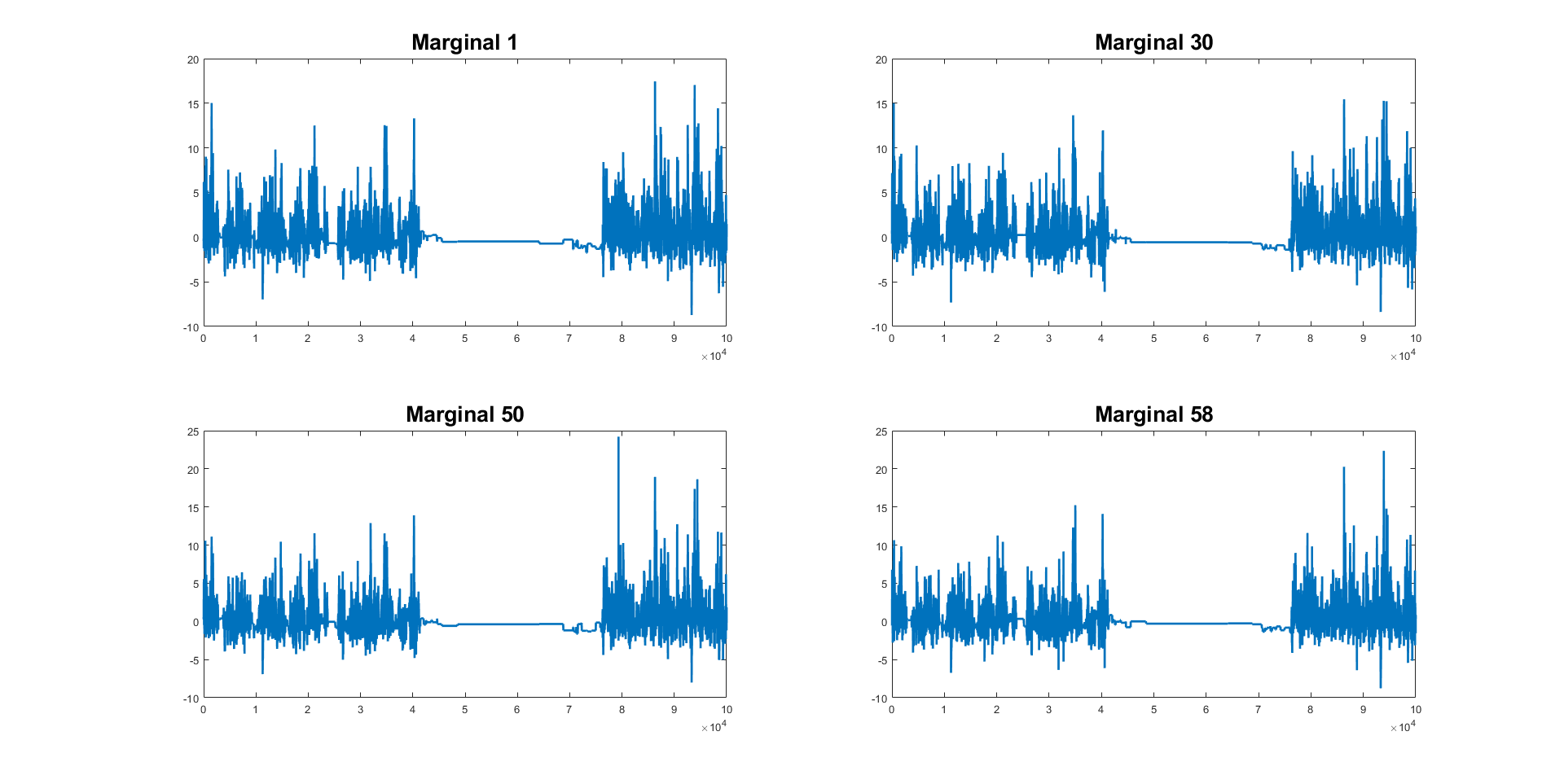}
\end{figure}

\begin{figure}[H]
\caption{The inefficiency factor (IACT) of the parameters of the skewed and heavy-tailed high-dimensional target distribution in section \ref{subsec:Multivariate t-distribution} estimated using HMC.
\label{fig:MCMC_TDIST_HMC_BOXPLOT}}
\centering{}\includegraphics[width=15cm,height=8cm]{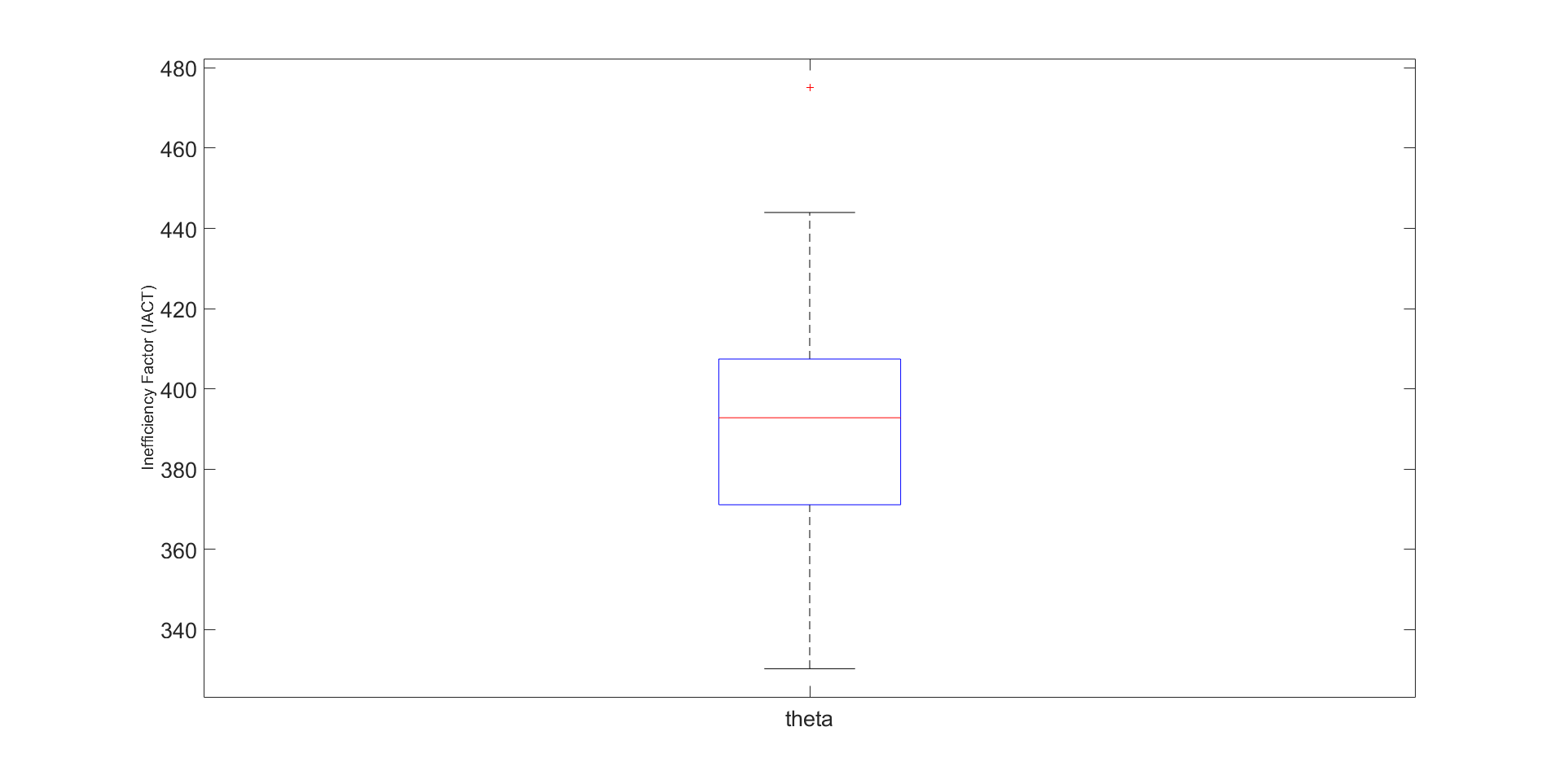}
\end{figure}

\section{Additional Figures for the Multimodal High-Dimensional Target Distribution Example \label{additionalfigure}}

This section gives additional figures for the multimodal high-dimensional target distribution example in section \ref{subsec:Multivariate-Mixture-of normals}.

\begin{figure}[H]
\caption{The plot of the average lower bound values over the last 500 steps for the variational approximations (A1)-(A6) for the 100-dimensional
mixture of normals example.\label{CopMixNomMixNomExample}}

\centering{}\includegraphics[width=15cm,height=6cm]{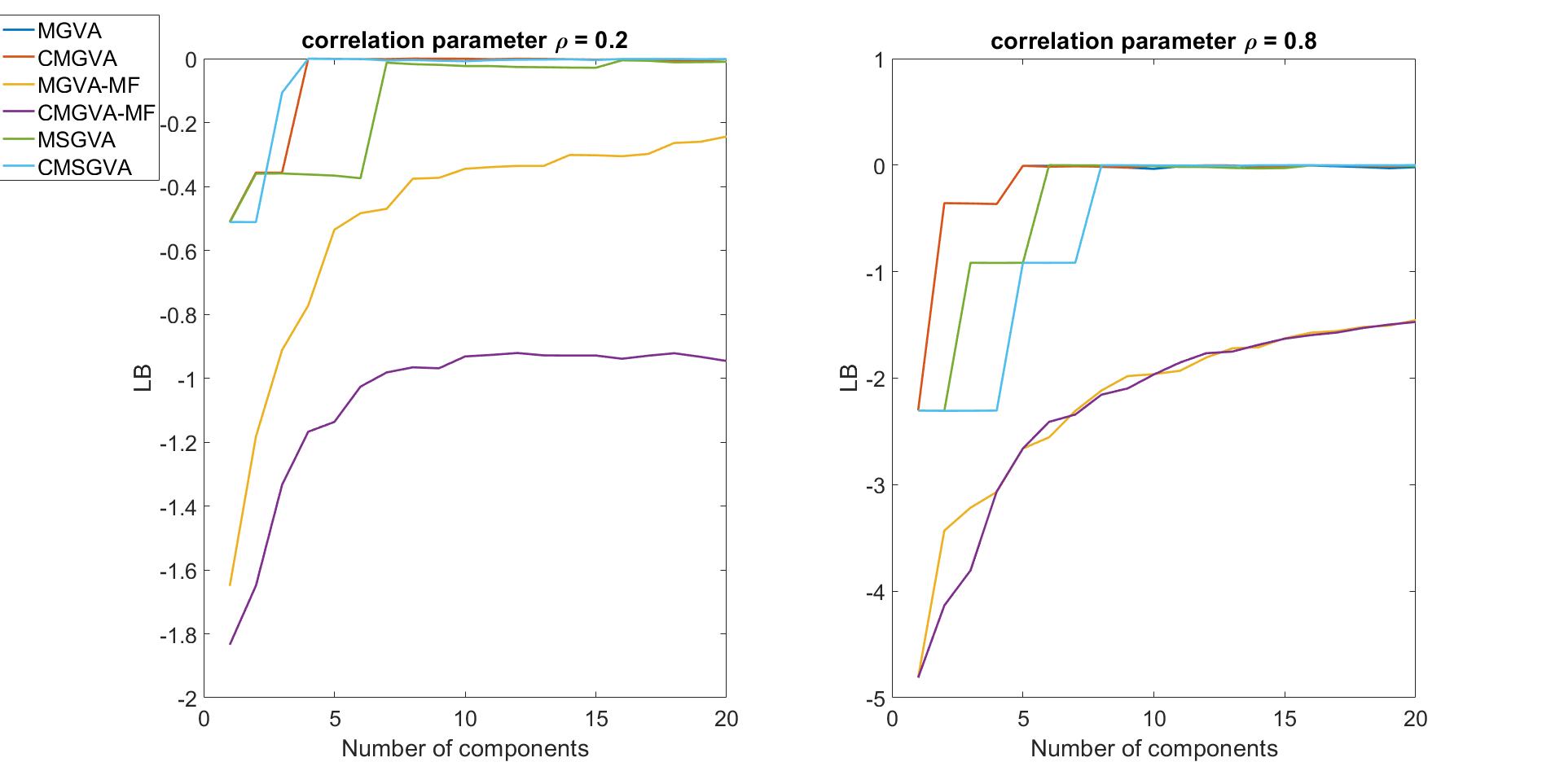}
\end{figure}

\begin{figure}[H]
\caption{Top panels: The plot of the YJ-parameters $\gamma_{i}$ for all $i=1,...,m$ of the 2-components CMGVA and 2-components CMSGVA for the mixture of normals example. Bottom panels: The plot of $\widetilde{\alpha_{i}}$ for all $i=1,...,m$ of the 2-components MSGVA and 2-components CMSGVA for the mixture of normals example.  \label{YJ_plot_mixnom}}

\centering{}\includegraphics[width=15cm,height=6cm]{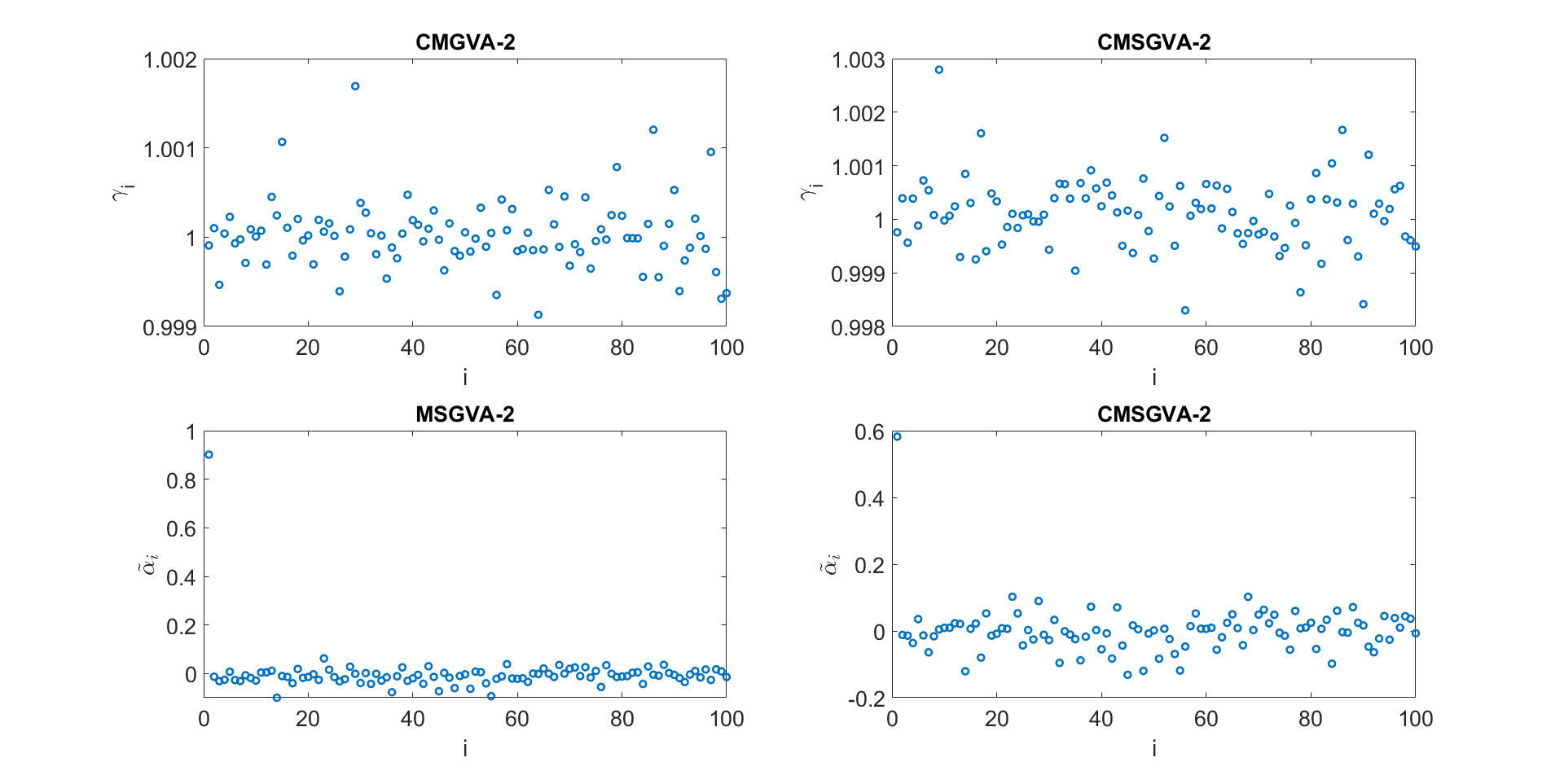}
\end{figure}

\begin{figure}[H]
\caption{Kernel density estimates of some of the marginal parameters $\theta$ approximated
with Gaussian, Gaussian Copula, skew Gaussian, skew Gaussian copula, and the optimal variational approximations (A1)-(A6) 
for the mixture of normals example with $\rho=0.2$ \label{fig:Kernel-Density-Estimates mixnormal mix100dim02}}

\centering{}\includegraphics[width=15cm,height=5cm]{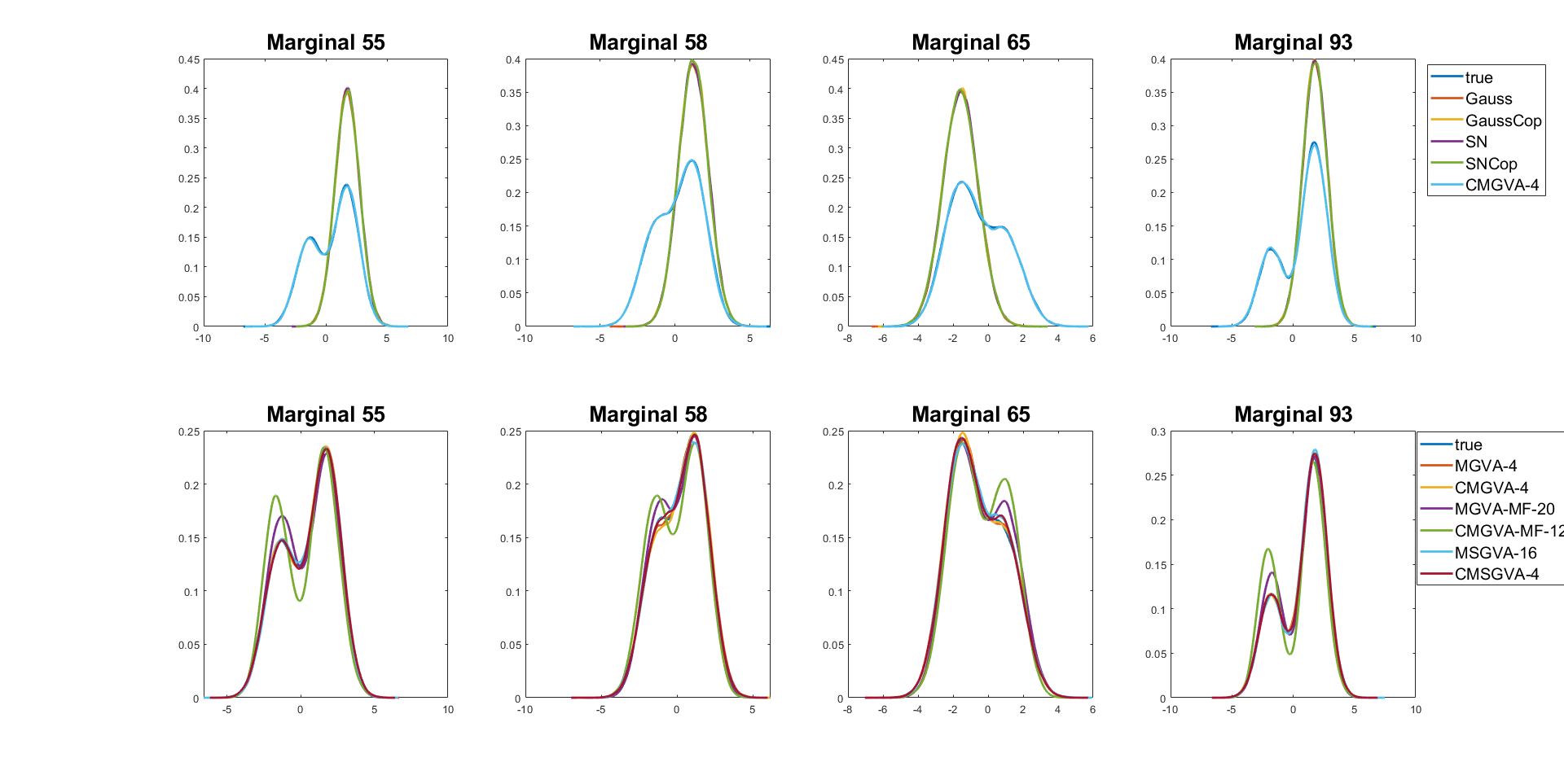}
\end{figure}

\begin{figure}[H]
\caption{Top: Scatter plot of the observations generated from true target distribution
(blue), and a 4-component CMGVA (orange) for the mixture of normals
example with $\rho=0.2$; Bottom: Scatter plot of the observations
generated from true target distribution (blue), and Gaussian copula (orange)
for the mixture of normals example with $\rho=0.2$ \label{fig:bivariatescatterplotmixnormal02}}

\centering{}\includegraphics[width=15cm,height=8cm]{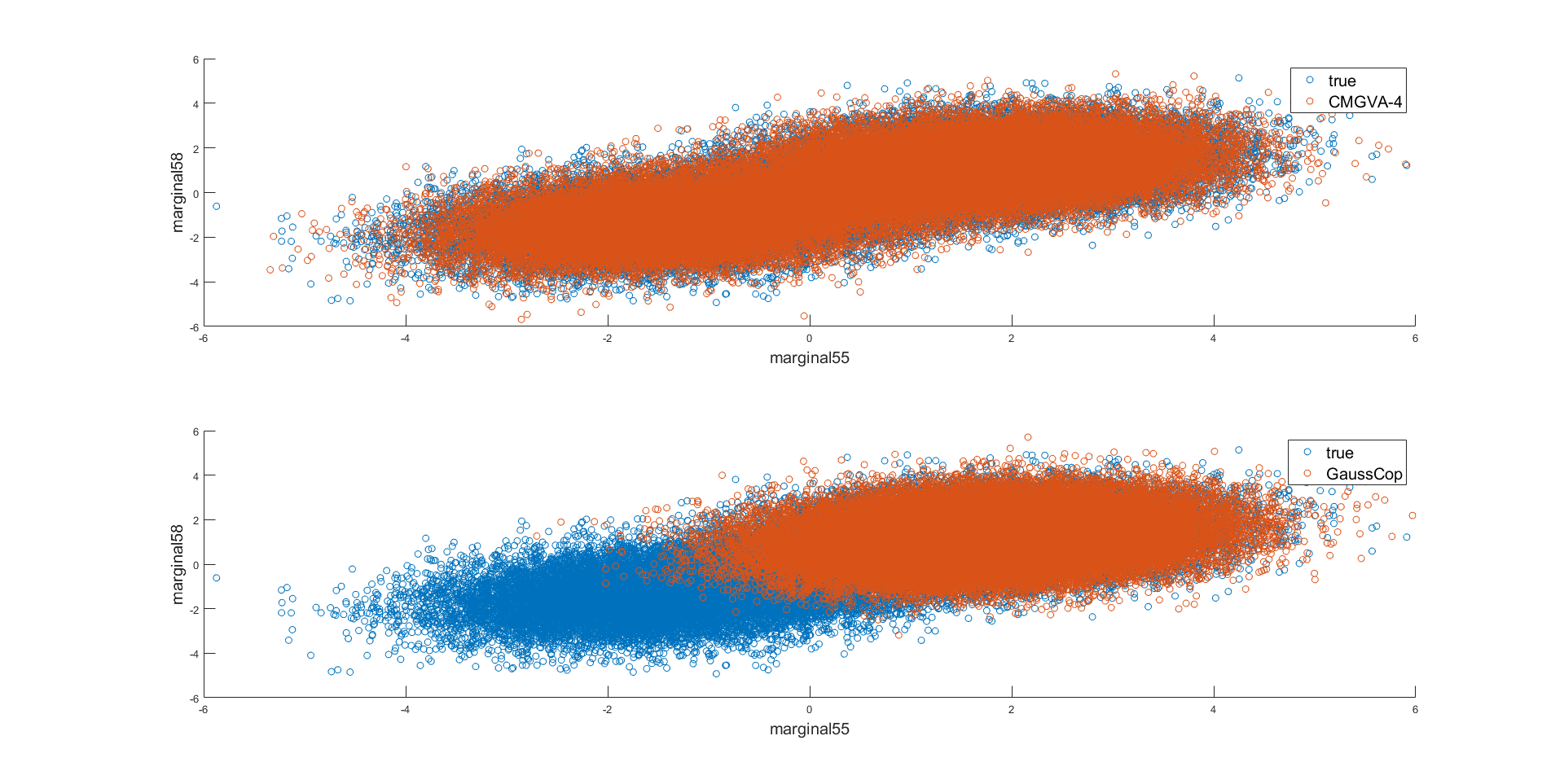}
\end{figure}

\begin{figure}[H]
\caption{The plot of the average lower bound values over the last 500 steps for the CMGVA for the 100-dimensional
normal distribution example.\label{CopNomExample}}

\centering{}\includegraphics[width=15cm,height=8cm]{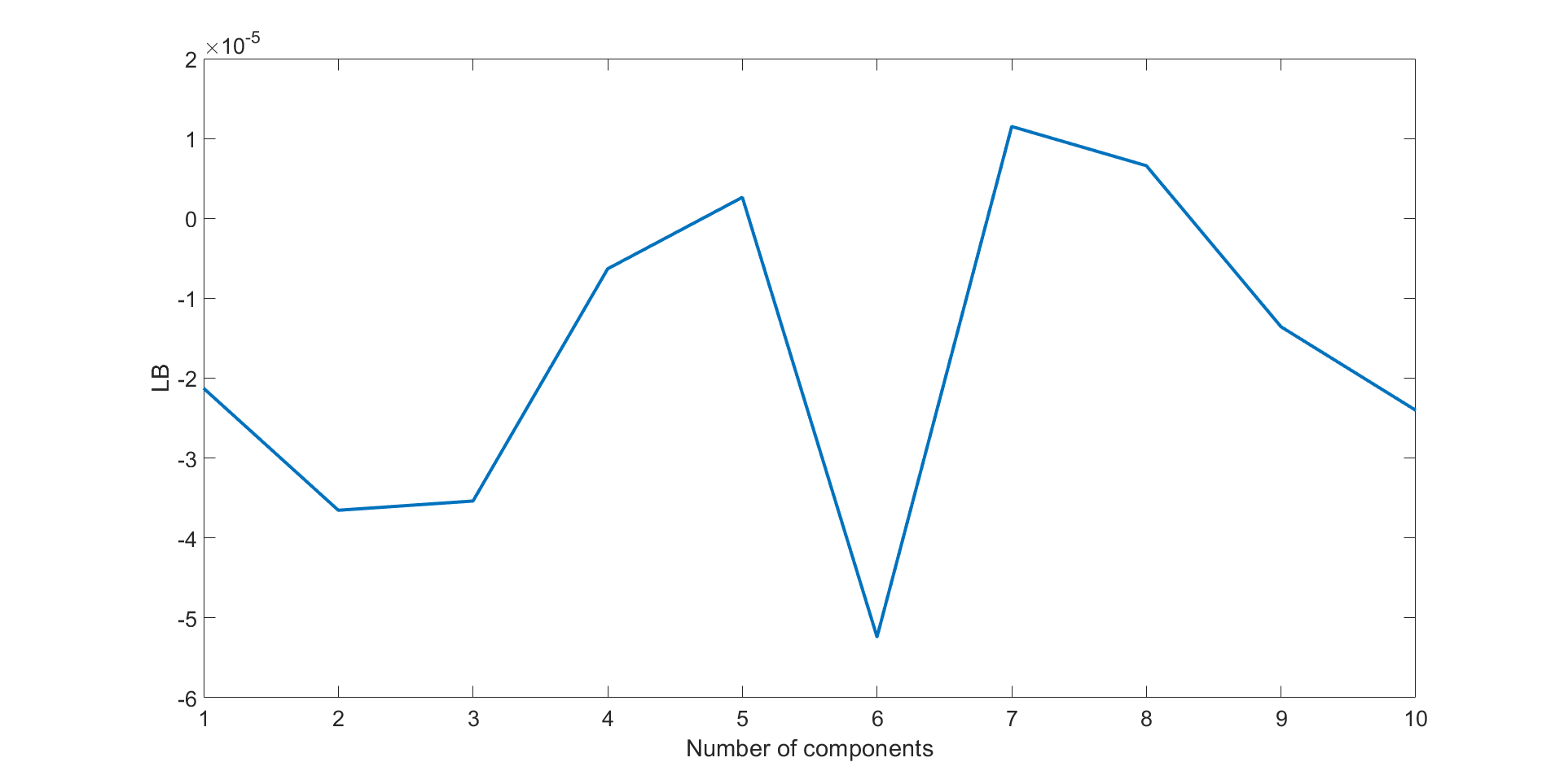}
\end{figure}

\begin{figure}[H]
\caption{The trace plots of some of the marginal distributions of the multimodal high-dimensional target distribution in section \ref{subsec:Multivariate-Mixture-of normals}.
\label{fig:MCMC_MIXNOM_HMC_TRACE}}
\centering{}\includegraphics[width=15cm,height=8cm]{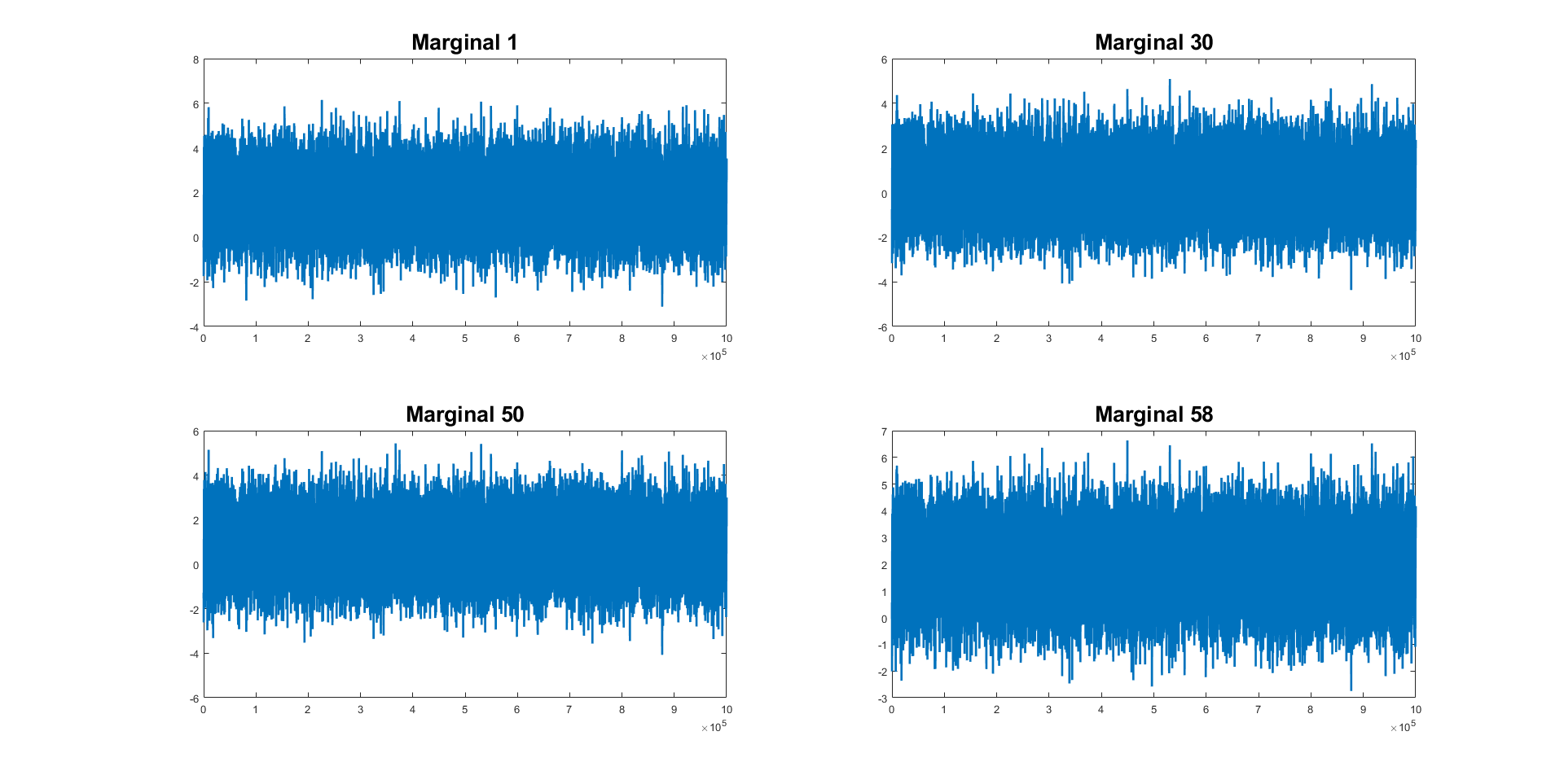}
\end{figure}

\begin{figure}[H]
\caption{The inefficiency factor (IACT) of the parameters of the multimodal high-dimensional target distribution in section \ref{subsec:Multivariate-Mixture-of normals} estimated using HMC.
\label{fig:MCMC_MIXNOM_HMC_BOXPLOT}}
\centering{}\includegraphics[width=15cm,height=8cm]{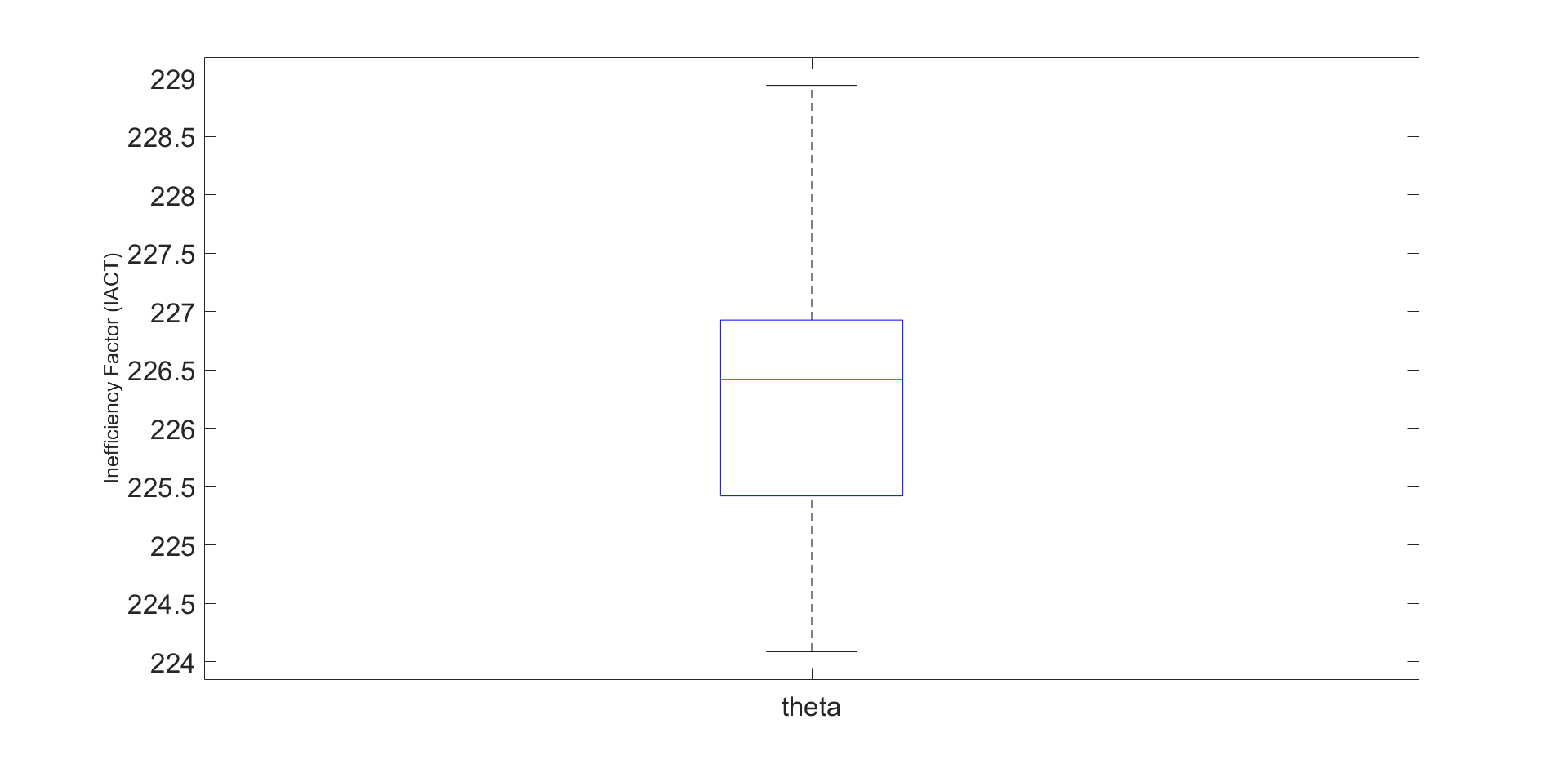}
\end{figure}


\section{Planar Flows\label{planarflows}}

This section discusses variational inference using planar
flows proposed
by \citet{rezende2015variational}. The planar flows consider a family of transformations of the
form
\[
f\left(\theta\right)=\theta+uh\left(w^{\top}\theta+b\right),
\]
where $\lambda=\left\{ w\in R^{D},u\in R^{D},b\in R\right\} $ are
the variational parameters and $h\left(\cdot\right)$ is a smooth
elementwise non-linearity with derivative $h^{'}\left(\cdot\right)$.
The log determinant of the Jacobian term is 

\begin{equation}
\log\left|\det\frac{\partial f\left(\theta\right)}{\partial\theta}\right|=\log\left|1+u^{\top}\psi\left(\theta\right)\right|,
\end{equation}
where $\psi\left(\theta\right)=h^{'}\left(w^{\top}\theta+b\right)w$.
The density of $q_{K}\left(\theta\right)$ is obtained by successively
transforming a random variable $\theta_{0}$ with distribution $q_{0}$
through a chain of $K$ transformations is 
\begin{eqnarray}
\theta_{K} & = & f_{K}\circ...\circ f_{2}\circ f_{1}\left(\theta_{0}\right),\\
\log q_{K}\left(\theta_{K}\right) & = & \log q_{0}\left(\theta_{0}\right)-\sum_{k=1}^{K}\log\left|1+u_{k}^{\top}\psi_{k}\left(\theta_{k-1}\right)\right|.
\end{eqnarray}
Section \ref{subsec:Bayesian-Deep-Neural} compares the proposed variational approximations with the
planar flows with flow lengths of $10$ transformations for
the neural nets with (8,5,5,1), (8,10,10,1), and (8,20,20,1) structures
for the auto dataset and neural nets with (9,5,5,1), (9,10,10,1),
and (9,20,20,1) structures for the abalone dataset. The non-linearity
function $h\left(\cdot\right)$ is the tanh function and the initial
distribution is a Gaussian distribution with a mean vector $\mu$ and a diagonal
covariance matrix with elements $d=\left(d_{1},...,d_{M}\right)$,
where $M$ is the number of parameters. We use $S=1000$ samples to accurately
estimate the gradients of the lower bound. The optimisation algorithm
is stopped if it exceeds $10000$ iterations or the lower bound does
not improve after $20$ iterations. To reduce the noise in estimating the lower bound, we take the average of the lower bound over a moving window of $250$ iterations \citep[see][for further details]{Tran2017}. We also use the momentum method \citep{polyak1964some} to help accelerating stochastic
gradient optimisation and reduce the noise in the estimated gradients
of the lower bound. 

\begin{figure}[H]
\caption{The plots of the lower bound values for planar flows with flow lengths
10  for a neural
net with a (8,20,20,1) structure for the auto dataset and a (9,20,20,1) structure
for the abalone dataset. \label{fig:The-plots-ofnn2020}}

\centering{}\includegraphics[width=15cm,height=10cm]{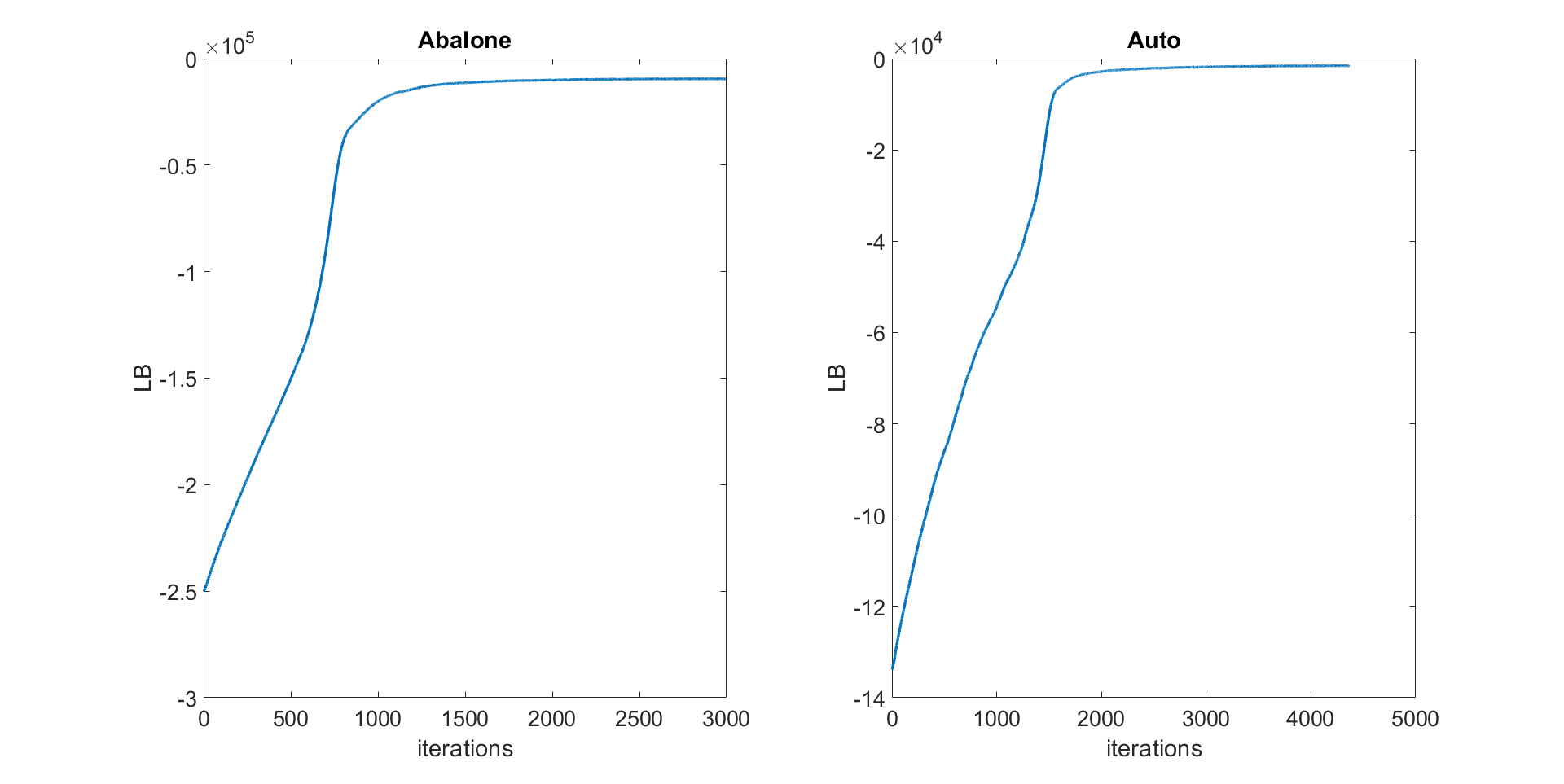}
\end{figure}

Figures
\ref{fig:The-plots-ofnn5} and \ref{fig:The-plots-ofnn2020} show that the lower bound of the planar flows for the two
datasets for neural nets with different structures increase at the
start and then converge. 


\begin{figure}[H]
\caption{The plots of the lower bound values for planar flows with flow lengths
10 for a neural
net with a (8,5,5,1) structure for the auto dataset and a (9,5,5,1) structure
for the abalone dataset. \label{fig:The-plots-ofnn5}}

\begin{centering}
\includegraphics[width=15cm,height=10cm]{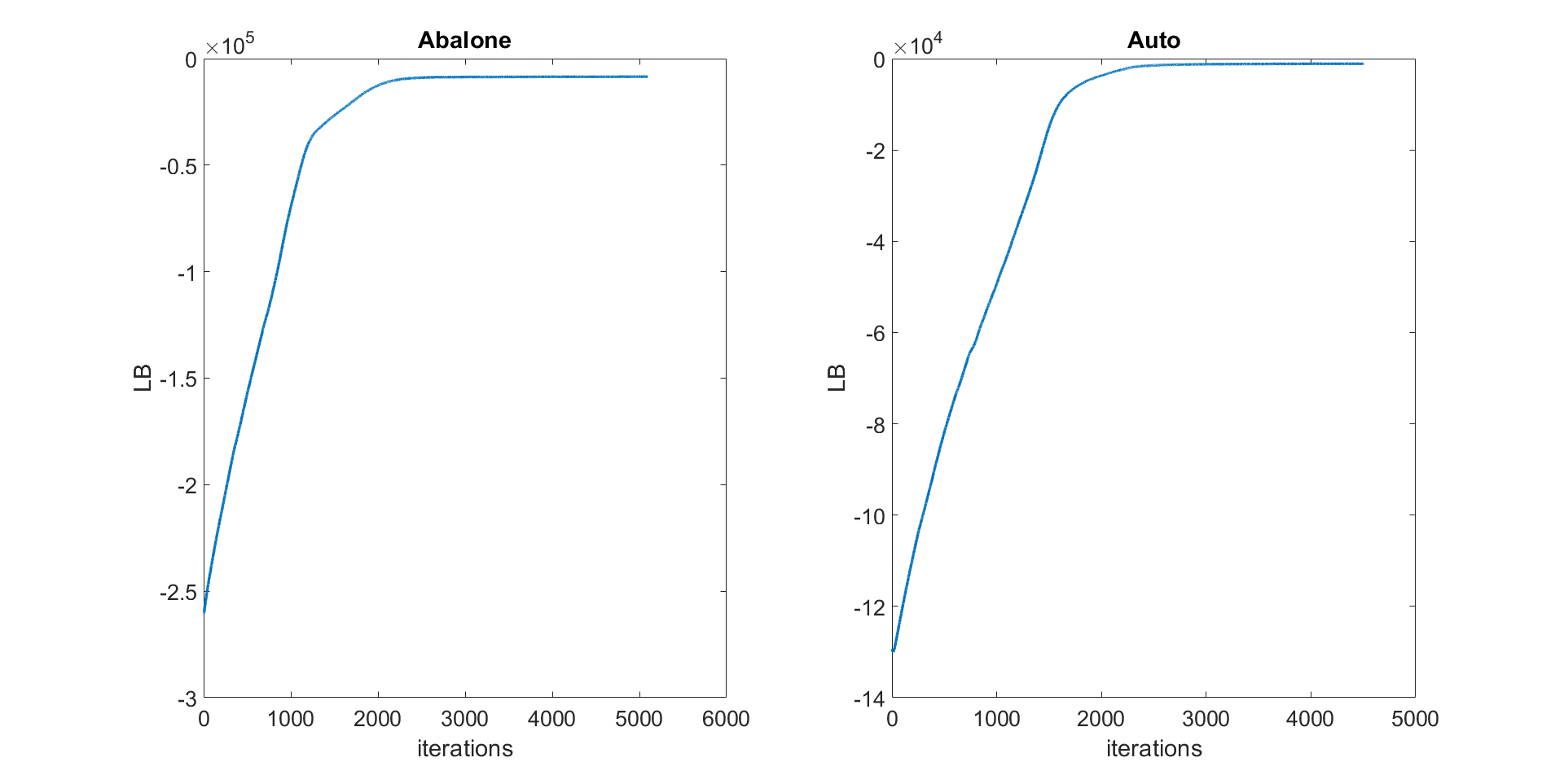}
\par\end{centering}
\end{figure}

\begin{figure}[H]
\caption{The plots of the lower bound values for planar flows with flow lengths
10 for a neural
nets with a (8,10,10,1) structure for the auto dataset and a (9,10,10,1) structure
for the abalone dataset. \label{fig:The-plots-ofnn5}}

\begin{centering}
\includegraphics[width=15cm,height=10cm]{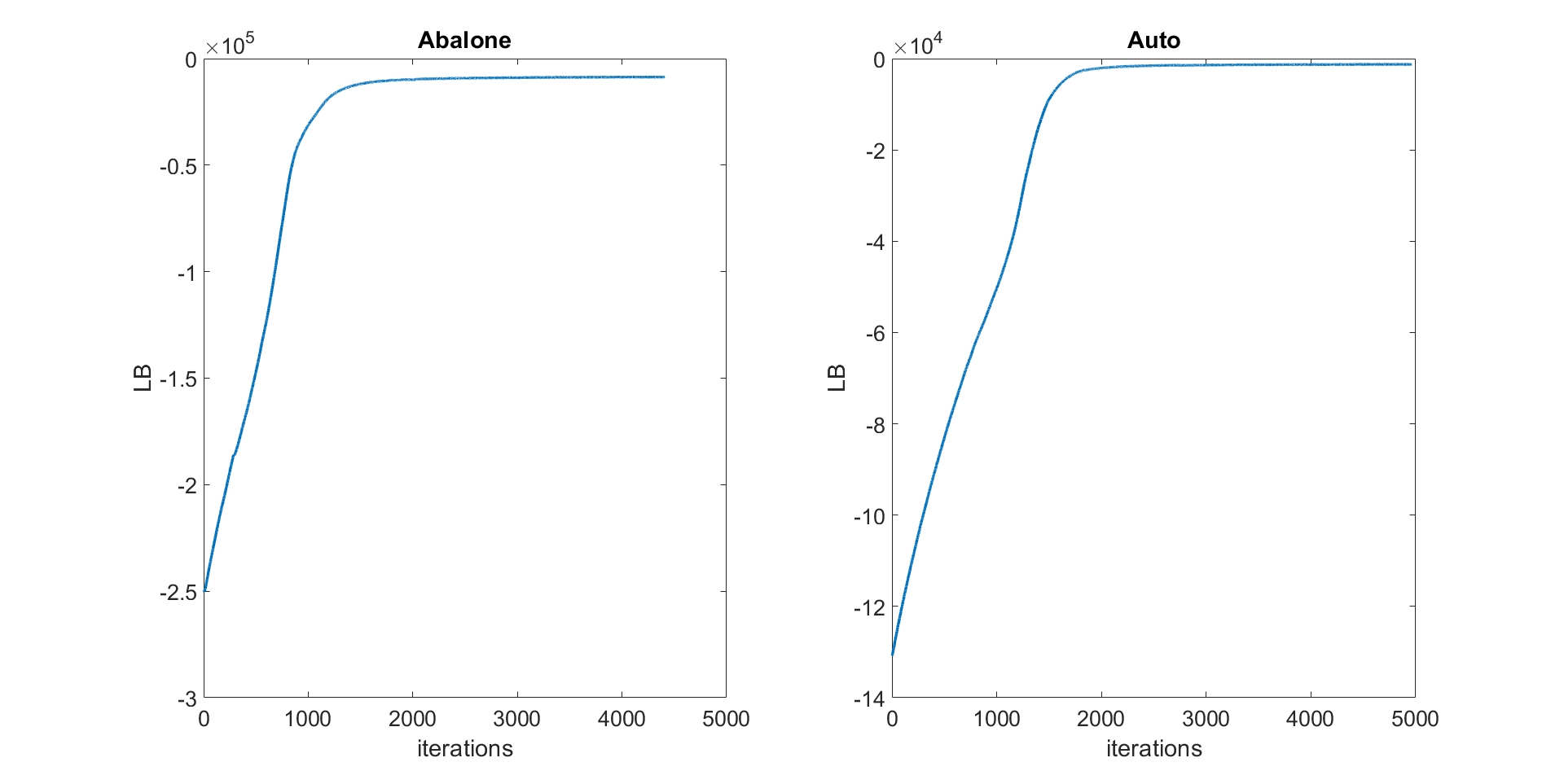}
\par\end{centering}
\end{figure}

\section{Stopping Criterion for the Flexible Bayesian Regression with a Deep Neural Network Example\label{stoppingNN}}
This section discusses the stopping criterion used for the optimisation algorithm in Section \ref{subsec:Bayesian-Deep-Neural}. The algorithm is stopped 
if it exceeds $P$ iterations or the lower bound does not improve after $20$ iterations.
For the first component, we set $P$ to $5000$ iterations and for the subsequent components, we set $P$ to $1000$ iterations, with the exception of the neural nets (8,20,20,1) for the auto dataset and (9,20,20,1) for the abalone dataset, where we set $P$ to $5000$ due to a greater number of parameters. 
To reduce the noise in estimating the lower bound, we average the lower bound over a moving window of $100$ iterations for the first component and $250$ for the additional components due to more challenging optimisation problems 
\citep[see][for further details]{Tran2017}. We also use the momentum method \citep{polyak1964some} to help accelerating stochastic
gradient optimisation and reduce the noise in the estimated gradients
of the lower bound for the first component.
The step sizes are set to the values given in section \ref{subsec:Learning-Rate} of the online supplement, except $\alpha_{\mu}$ is set to 0.001.


\section{Additional Figures and Tables for the Flexible Bayesian Regression with a Deep Neural Network Example\label{additionalfiguresNN}}

\begin{figure}[H]
\caption{The trace plots of some of the parameters of the (9,10,10,1) neural net structure for the auto dataset.
\label{fig:MCMC_NN_ABALONE}}
\centering{}\includegraphics[width=15cm,height=8cm]{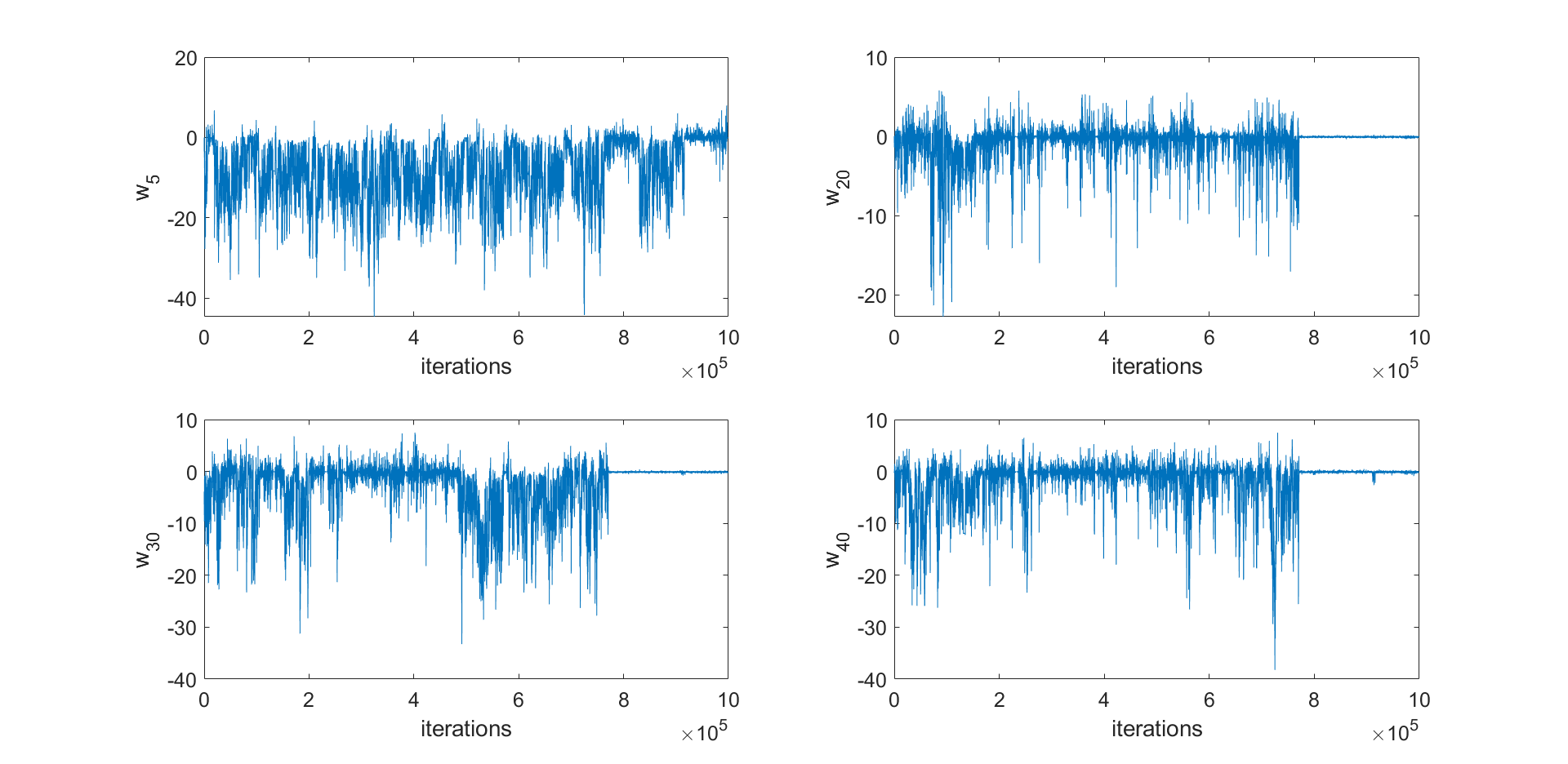}
\end{figure}

Figure \ref{fig:MCMC_NN_ABALONE} shows the trace plots of the parameters of the neural net with the (9,10,10,1) structure for the auto dataset. The parameters do not show evidence of convergence even after $1000000$ iterations.

\begin{figure}[H]
\caption{The inefficiency factor (IACT) of the parameters of the of the neural net with the (9,10,10,1) structure for the auto dataset estimated using HMC.
\label{fig:MCMC_NN_HMC_BOXPLOT}}
\centering{}\includegraphics[width=15cm,height=8cm]{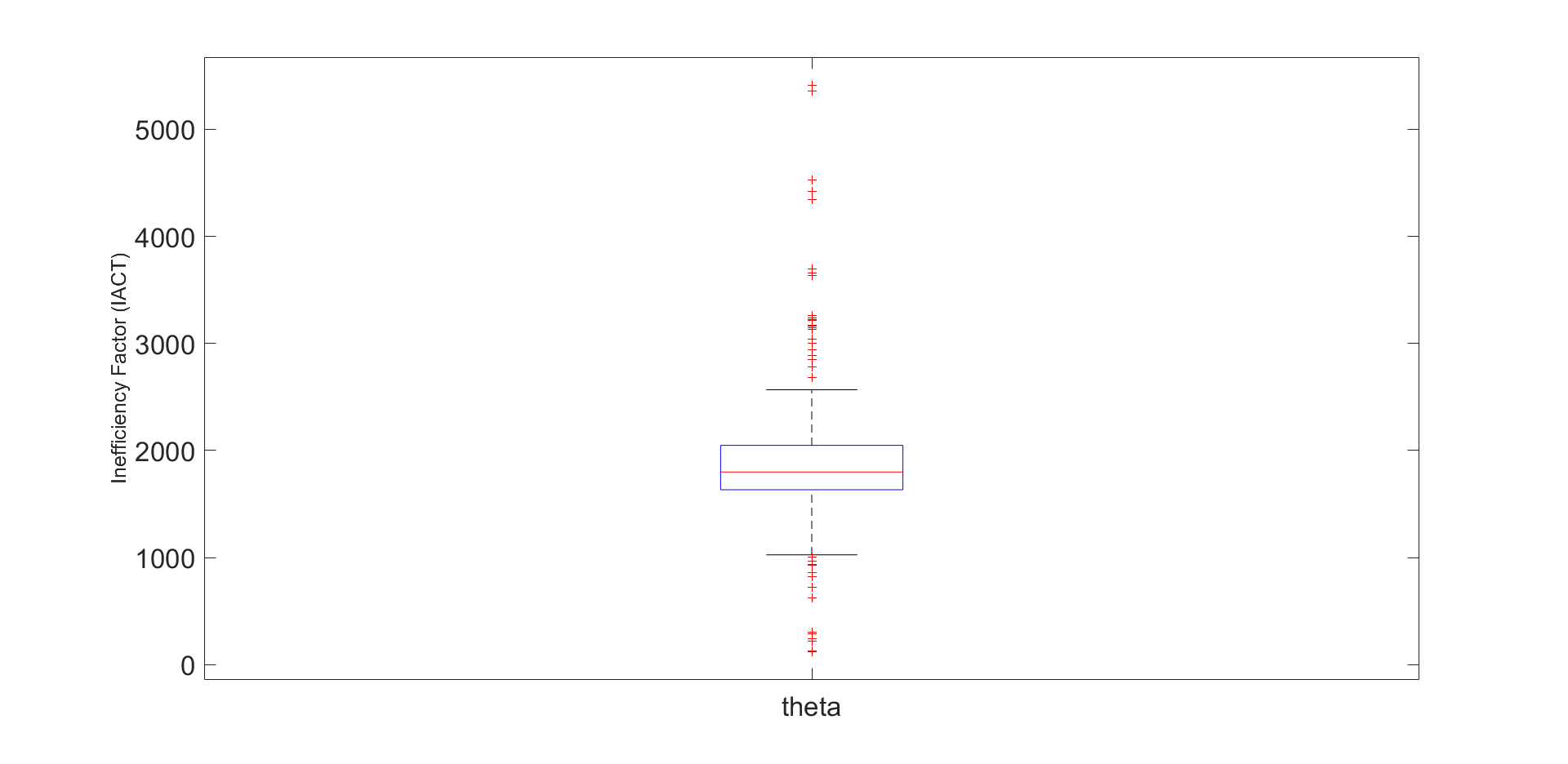}
\end{figure}

\begin{figure}[H]
\caption{Top panels: Plots of the average lower bound values over the last 100 steps for the variational approximations A1 and A2 for the (8,5,5,1) neural net structure for the auto dataset and the (9,5,5,1) neural net structure for the abalone dataset. Bottom panels: Plots of the log of the estimated approximate posterior predictive scores for the variational approximations A1 and A2 for the (8,5,5,1) neural net structure for the auto dataset and the (9,5,5,1) neural net structure for the abalone dataset.
\label{fig:LB_pred_55}}

\centering{}\includegraphics[width=15cm,height=8cm]{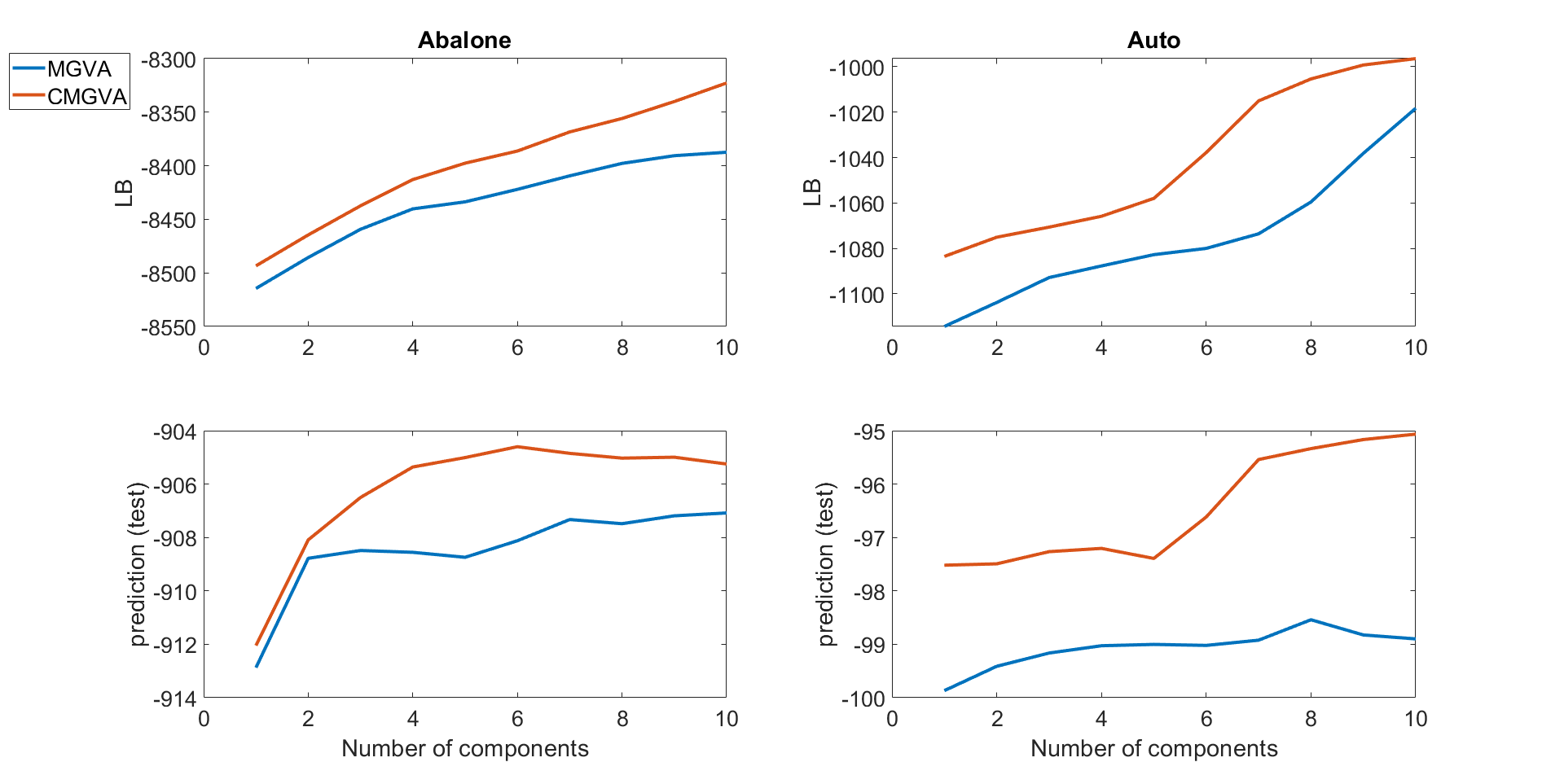}
\end{figure}

\begin{table}[H]
\caption{Comparing the performance of optimal variational approximation A2 (with the number in bracket indicating the optimal number of components)
with the planar flow with flow lengths of $10$ transformations in terms of lower bound (top panel) and predictive
values (bottom panel).\label{CMGVAagainstPlanar}}

\centering{}%
\begin{tabular}{cccc}
\hline 
Data & Neural Nets & Planar & Optimal A2\tabularnewline
\hline 
Abalone & $\left(9,5,5,1\right)$ & $-8495.38$ & $-8322.75(10)$\tabularnewline
 & $\left(9,10,10,1\right)$ & $-8657.56$ & $-8492.98(10)$\tabularnewline
 & $\left(9,20,20,1\right)$ & $-9481.93$ & $ -8786.62(10)$\tabularnewline
Auto & $\left(8,5,5,1\right)$ & $-1127.39$ & $-996.41(10)$\tabularnewline
 & $\left(8,10,10,1\right)$ & $-1255.76$ & $-1092.79(10)$\tabularnewline
 & $\left(8,20,20,1\right)$ & $-1560.68$ & $ -1389.68(10)$\tabularnewline
\hline 
Abalone & $\left(9,5,5,1\right)$ & $-914.80$ & $-904.60(6)$\tabularnewline
 & $\left(9,10,10,1\right)$ & $-912.97$ & $ -893.18(10) $\tabularnewline
 & $\left(9,20,20,1\right)$ & $-920.56$ & $ -906.51(6) $\tabularnewline
Auto & $\left(8,5,5,1\right)$ & $-98.05$ & $-95.06(10)$\tabularnewline
 & $\left(8,10,10,1\right)$ & $-98.09$ & $-95.19(9)$\tabularnewline
 & $\left(8,20,20,1\right)$ & $-98.08$ & $ -95.30(8)$\tabularnewline
\hline 
\end{tabular}
\end{table}

\bibliographystyle{apalike}
\bibliography{references_v1}
\end{document}